\documentclass[iop]{emulateapj}

 \usepackage{graphicx}
 \usepackage{natbib}
 \usepackage[usenames,dvipsnames]{color}
 \usepackage{enumitem}
 \usepackage[T1]{fontenc}
 \usepackage{apjfonts}
 \usepackage{amssymb,amsmath,amsbsy}
 \usepackage{microtype}
 \usepackage[nolist,nohyperlinks]{acronym}
 \usepackage[breaklinks,colorlinks,citecolor=blue,bookmarksnumbered,
             bookmarksopen,hyperfootnotes=false]{hyperref}

  \newcommand{\Tbol}  {T_\mathrm{bol}}
  \newcommand{\Lbol}  {L_\mathrm{bol}}
  \newcommand{\Lcm}   {L_\mathrm{cm}}
  \newcommand{\kms}   {km~s$^{-1}$}
  
  \newcommand{\jpb}   {Jy~beam$^{-1}$}    
  \newcommand{\lo}    {$L_{\sun}$}
  \newcommand{\mj}    {M$_\textrm{J}$}
  \newcommand{\mo}    {$M_{\sun}$}
  \newcommand{\moyr}  {$M_{\sun}$ yr$^{-1}$}
  \newcommand{\mdotout} {\dot{M}_\mathrm{out}}
  \newcommand{\co}    {$^{12}$CO}
  \newcommand{\eg}    {e.\,g.,}
  \newcommand{\ie}    {i.\,e.,}
  \newcommand{\tnmk}[1] {\,$^{#1}$}
  \newcommand{\tntxt}[2] {~\,\,$^{#1}$#2}
  \newcommand{\hang} {\hangindent=1em\hangafter=1}

  \newcommand{\linss} {\noalign{\smallskip}\hline\noalign{\smallskip}}

\begin{document}

\title{First detection of thermal radio jets in a sample of proto-brown
  dwarf candidates}

 \shorttitle{First detection of thermal radio jets in proto-BD candidates}

\author{
  Oscar Morata\altaffilmark{1}, 
  Aina Palau\altaffilmark{2}, 
  Ricardo F.~Gonz\'alez\altaffilmark{2},
  Itziar de Gregorio-Monsalvo\altaffilmark{3,4},
  \'Alvaro Ribas\altaffilmark{5,6,7},
  Manuel Perger\altaffilmark{8},
  Herv\'e Bouy\altaffilmark{6},
  David Barrado\altaffilmark{6},
  Carlos Eiroa\altaffilmark{9},
  Amelia Bayo\altaffilmark{10,11},
  Nuria Hu\'elamo\altaffilmark{6},
  Mar\'{i}a Morales-Calder\'on\altaffilmark{6},
  and Lu\'{i}s F. Rodr\'{i}guez\altaffilmark{2}
  }

  \altaffiltext{1}{Institute of Astronomy and Astrophysics, Academia Sinica,
    P.O.\ Box 23-141, Taipei 106, Taiwan}
  \altaffiltext{2}{Centro de Radioastronom\'{i}a y Astrof\'{i}sica, Universidad
    Nacional Aut\'onoma de M\'exico, P.O.~Box 3-72, 58090 Morelia,
    Michoac\'an, M\'exico}
  \altaffiltext{3}{Joint ALMA Observatory (JAO), Alonso de C\'ordova 3107,
    Vitacura, Santiago, Chile}
  \altaffiltext{4}{European Southern Observatory, Karl Schwarzschild Str 2,
    85748, Garching bei M\"unchen, Germany}
  \altaffiltext{5}{European Space Astronomy Centre (ESA), PO Box 78, 28691
    Villanueva de la Ca\~nada, Madrid, Spain}
  \altaffiltext{6}{Centro de Astrobiolog\'{i}a, INTA-CSIC,
    Dpto.\ Astrof\'{i}sica, ESAC Campus, P.O.~Box 78, 28691 Villanueva de la
    Ca\~nada, Madrid, Spain}
  \altaffiltext{7}{Ingenier\'{i}a y Servicios Aeroespaciales-ESAC, PO Box 78,
    28691 Villanueva de la Ca\~nada, Madrid, Spain} 
  \altaffiltext{8}{Institut de Ci\`encies de l'Espai (CSIC-IEEC), Campus UAB
  --  Facultat de Ci\`encies, Torre C5 -- parell 2, E-08193 Bellaterra,
  Catalunya, Spain} 
  \altaffiltext{9}{Departamento de F\'{i}sica Te\'orica, Facultad de Ciencias,
    Universidad Aut\'onoma de Madrid, Cantoblanco, E-28049 Madrid, Spain}
  \altaffiltext{10}{Max Planck Institut f\"ur Astronomie, K\"onigstuhl 17,
    D-69117, Heidelberg, Germany}
  \altaffiltext{11}{Departamento de F\'{i}sica y Astronom\'{i}a, Facultad de
    Ciencias, Universidad de Valpara\'{i}so, Av. Gran Breta\~na 1111, 5030
    Casilla, Valpara\'{i}so, Chile}

  \email{omorata@asiaa.sinica.edu.tw}

 \shortauthors{Morata et al.}

\begin{abstract}

  We observed with the \acl{JVLA} at 3.6 and 1.3 cm a sample of 11 proto-brown
  dwarf candidates in Taurus in a search for thermal radio jets driven by the
  most embedded \aclp{BD}.
  We detected for the first time four thermal radio jets in proto-brown dwarf
  candidates.
  We compiled data from UKIDSS, 2MASS, \emph{Spitzer}, WISE and
  \emph{Herschel} to build the \ac{SED} of the objects in our sample, which
  are similar to typical Class~I \acp{SED} of \acp{YSO}.
  The four proto-brown dwarf candidates driving thermal radio jets also roughly
  follow the well-known trend of centimeter luminosity against bolometric
  luminosity determined for \acp{YSO}, assuming they belong to Taurus,
  although they present some excess of radio emission compared to the known
  relation for \acp{YSO}.
  Nonetheless, we are able to reproduce the flux densities of the radio jets
  modeling the centimeter emission of the thermal radio jets using the same
  type of models applied to \acp{YSO}, but with corresponding smaller stellar
  wind velocities and mass-loss rates, and exploring different possible
  geometries of the wind or outflow from the star.
  Moreover, we also find that the modeled mass outflow rates for the
  bolometric luminosities of our objects agree reasonably well with the trends
  found between the mass outflow rates and bolometric luminosities of
  \acp{YSO}, which indicates that, despite the ``excess'' centimeter emission,
  the intrinsic properties of proto-brown dwarfs are consistent with a
  continuation of those of very low mass stars to a lower mass range.
  Overall, our study favors the formation of brown dwarfs as a scaled-down
  version of low-mass stars.
  \acresetall

\end{abstract}


\section{Introduction}

  A crucial question that has been at the core of recent vigorous discussions
  is that of the formation of \acp{BD}.
  It is generally accepted that \acp{BD} form by gravitational instability of
  a very low-mass dense core, on a dynamical timescale and with initial
  elemental composition similar to low-mass stars, as opposed to planet
  formation, which could happen by aggregation of a rocky core from smaller
  planetesimals, on timescales longer than a dynamical time, and with
  elemental composition with an overall deficit on light elements
  \citep{Whitworth2007}.
  On the other hand, the underlying mechanism responsible for the formation of
  the very low-mass dense cores that would form \acp{BD} is not clear yet, and
  several scenarios were proposed to interpret the different observational
  results \citep[see
    \eg][]{Reipurth2001,Kroupa2003,Umbreit2005,Whitworth2007,Andre2012}.
  After a major theoretical and observational effort during the last decade,
  statistical studies of low mass star forming regions essentially comparing
  the properties of low-mass young stars and \acp{BD} in the Class~II/III
  stages (\ie\ well after the main accretion phase, \eg\ \citealt{Andre93})
  suggest that the dominant mechanism of \ac{BD} formation is
  indistinguishable from that of low-mass stars \citep[see
    \eg][]{Chabrier2014,Bayo11,Bayo12,Scholz12,AlvesOliveira13,Muzic14}.
  This is also favored by hydrodynamical simulations that routinely form
  \acp{BD} as a result of molecular cloud evolution, simultaneously
  reproducing the observed ratio of \acp{BD} to stars and the observed
  \ac{IMF} \citep[\eg][]{Bate12}.
  Thus, it seems that the dominant formation mechanism of \acp{BD} cannot be
  easily distinguished from that of low-mass stars, and the most promising
  mechanism is the fragmentation of turbulent clouds, which naturally form
  very low-mass dense cores due to the effects of turbulence \citep[see][for
    reviews]{Luhman2012,Chabrier2014}.
  However, up to now the turbulent fragmentation scenario is not yet directly
  supported by observations of deeply embedded \acp{BD}, what we call here
  `proto-\acp{BD}', \ie\ \acp{BD} in the stage equivalent to the Class~0/I
  stage of low-mass \acp{YSO} \citep{Andre93}, and the rest of the competing
  scenarios, mainly based on the halting of accretion of matter through
  ejection of protostellar embryos or disc fragments, and/or photo-erosion of
  pre-stellar cores, could still be possible
  \citep[\eg][]{StamatellosWhitworth09,Bate12,Luhman2012}.
  In fact, there are only two cases in the literature of Class~0/I
  proto-\acp{BD} \citep{Lee13,Palau2014}, and further candidates are
  definitely needed in order to compare in a statistically significant base
  the properties of proto-BDs to the properties of low-mass \acp{YSO}.
  Since \acp{BD} are expected to form as a scaled-down version of low-mass
  stars in the the `turbulent fragmentation' scenario, studying the properties
  of \acp{BD} in their most embedded phases of formation, and comparing their
  properties to the well-known relations established for low-mass protostars,
  should shed light on the formation mechanism of BDs.

 \begin{figure*}[t]
   \centering
   \includegraphics[width=\textwidth]{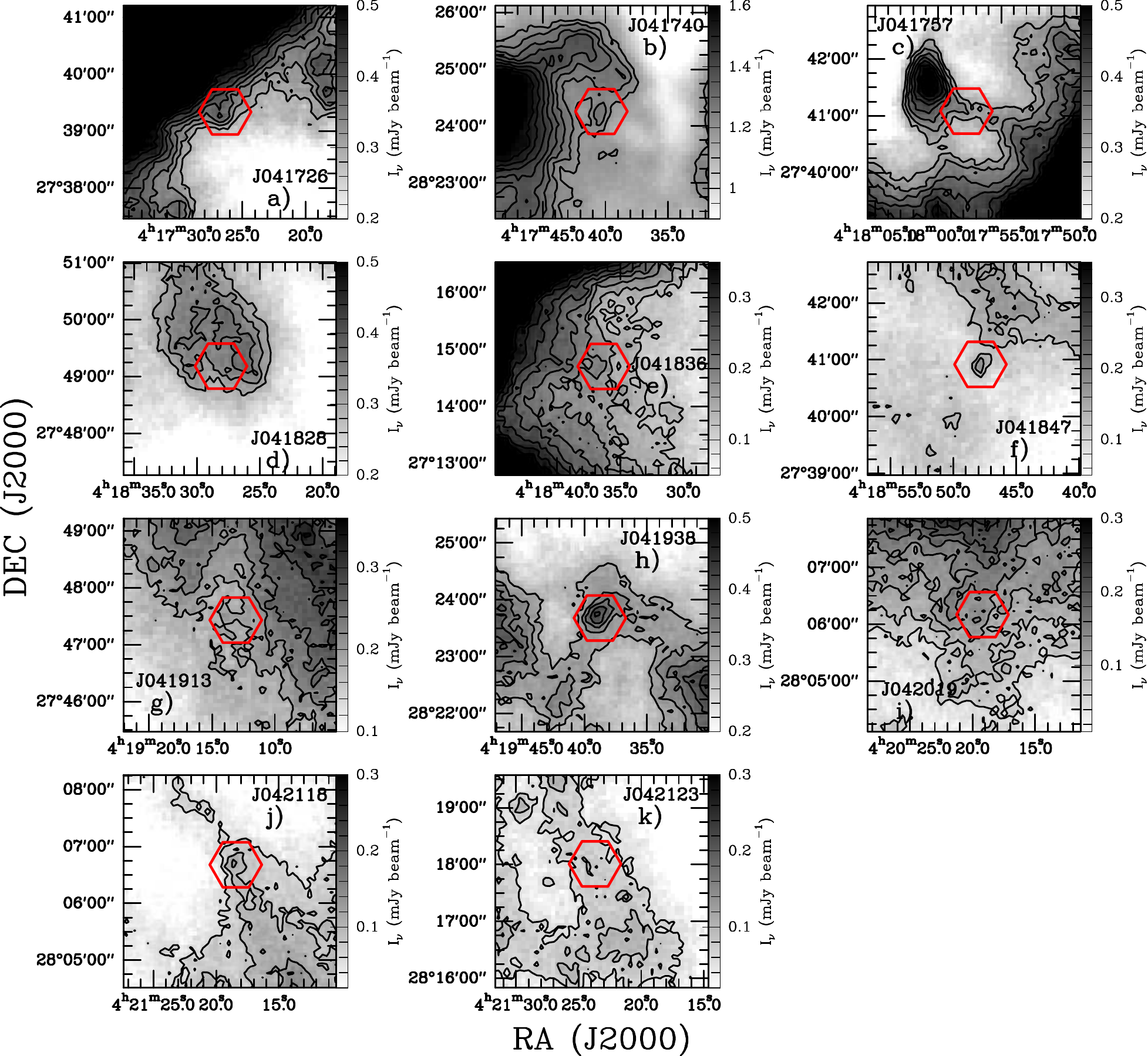}

   \caption{\emph{Herschel} SPIRE 250~$\mu$m continuum emission maps centered
     at the position of the proto-\ac{BD} candidates in Taurus, indicated
     by red hexagons, that we observed with the \ac{JVLA}.
     Contours are: \textit{a)} 0.3, 0.33, 0.36, 0.39, 0.42, 0.45, 0.48, 0.51
     m\jpb; \textit{b)} 1.20, 1.25, 1.30, 1.35, 1.40, 1.45, 1.50, 1.55, 1.60
     m\jpb; \textit{c)} 0.30, 0.34, 0.37, 0.40, 0.44, 0.47, 0.50; \textit{d)}
     0.30, 0.33. 0.36 m\jpb; \textit{e)} 0.13, 0.16, 0.19, 0.22, 0.25, 0.28,
     0.31, 0.34, 0.37, 0.40 m\jpb; \textit{f)} 0.13, 0.16, 0.19 m\jpb;
     \textit{g)} 0.19, 0.22, 0.27, 0.32, 0.37 m\jpb; \textit{h)} 0.30, 0.33,
     0.36, 0.39, 0.42 m\jpb; \textit{i)} 0.06, 0.09, 0.12, 0.14, 0.17,
     0.20, 0.23 m\jpb; \textit{j)} 0.06, 0.09, 0.12 m\jpb; and \textit{k)}
     0.06, 0.09 m\jpb, respectively. 
     }

 \label{fig:herschel}
 \end{figure*}

  The production of accretion-powered collimated ejections from the central
  protostellar object and disk is one of the processes characterizing the
  earliest phases of the evolution of high-, low-, and very-low mass \acp{YSO}
  \citep{Lada85,Shepherd1996,Li2014}.
  These mass ejections can be usually traced in different ways: as narrow,
  highly-collimated jets of atomic and/or molecular gas, with
  $v\sim100-1000$~\kms, observed from X-rays to mid-IR lines
  \citep{Bally2007,Ray2007,Frank2014}; or as less collimated but more massive
  molecular outflows with $v\sim1-30$~\kms, typically observed using molecular
  gas tracers like CO, HCO$^+$ or SiO at submillimeter/millimeter wavelengths.
  Another signpost of the ejection process is the presence of thermal radio
  jets, whose shock-ionized hydrogen atoms emit in the centimeter range with
  flat/positive spectral indices
  \citep[\eg][]{Rodriguez98,Beltran01,Reipurth02}.
  The centimeter emission of thermal radio jets in low-mass stars is thought
  to be free-free radiation produced by material (partially) ionized by the
  shock of the stellar wind with the surrounding gas \citep[see
    \eg][]{Anglada95,Curiel87}.
  Evidence for the connection between thermal radio jets and the wind from
  \acp{YSO} comes from the well-known trends between the centimeter luminosity
  and bolometric luminosity on one hand, and the centimeter luminosity and the
  momentum rate of the outflow on the other
  \citep[\eg][]{Anglada95,Bontemps96,Shirley07,AMI2011a,PhanBao14}.

  Several spectroastrometrically detected jets in \acp{BD} have been found in
  recent years \citep[see
    \eg][]{Whelan05,Whelan12,Joergens12,Joergens13,Riaz2015}, and a few
  molecular outflows have been detected, and imaged, in \acp{BD} or
  proto-\ac{BD} candidates at submillimeter/millimeter wavelengths
  \citep{PhanBao2008,PhanBao2014a,Palau2014,Monin2013}.
  However, only very few \acp{VeLLO} and proto-\ac{BD} candidates have been
  studied and detected in the centimeter range
  \citep{Andre99,Shirley07,Palau12}, making it difficult to test whether or
  not proto-\acp{BD} follow the trend between centimeter luminosity and
  bolometric luminosity.

  In this work, we present the results of the first search for thermal
  radio jets in a sample of proto-BD candidates.
  Sources were chosen from the sample of 12 proto-BD candidates selected by
  \citet{Barrado09} and \citet{Palau12} from \emph{Spitzer} color--color and
  color--magnitude diagrams.

  All these sources have red infrared colors and two of them (J041757 and
  J042118) were observed and detected at 350~$\mu$m with the \ac{CSO} 10-m
  telescope, where we detected two (small) dust condensations associated with
  the \emph{Spitzer} objects and condensations visible in Herschel maps at 160
  and 250 $\mu$m, respectively.
  We derived a mass for the gas traced by the dust emission of 1--10~\mj\ and
  0.3-3~\mj\ for J041757 and J042118, respectively.
  We also detected emission in J041757 at 3.6 and 6 cm in two VLA
  configurations, with a spectral index indicative of free-free thermal
  emission \citep{Palau12}.
  Unfortunately, problems with the calibration in the VLA-B configuration
  prevented us from having a good flux calibration for the resolved emission.
  Nonetheless, the source was an excellent candidate to have emission from a
  thermal radio jet, which was only pending to be confirmed with new
  observations.
  Additionally, we found an excess in blue-shifted emission, possibly
  indicating the blue wing of an outflow, around the position of J041757 in
  the IRAM 30-m spectra of the \co\ (1--0) line.
  The \acp{SED} that we could derive from the numerous multi-wavelength
  observations that our group had obtained \citep{Barrado09,Palau12} allowed
  us to derive for our sources bolometric temperatures, $\Tbol$, of 150--280~K
  and 140 ~K, respectively, typical of Class~I objects, and bolometric
  luminosities, $\Lbol$, $\lesssim0.005$ and $<0.0023$ \lo, respectively,
  which would place them in the proto-BD regime.
  \citet{Palau12} discussed the different possible types of objects that could
  be expected to explain the above observational data, but only the scenario
  of proto-\acp{BD} belonging to the Taurus Molecular Cloud consistently
  explains the properties of the emission ranging from optical, through IR to
  sub-mm wavelengths.

  The structure of this paper is as follows: we describe the selected sample
  and the observations carried out with the \ac{JVLA} in \autoref{sec:obs}.
  \autoref{sec:results} contains the results at 1.3 and 3.6 cm, the resulting
  spectral indices and the calculated \acp{SED} of the objects of our sample.
  We discuss the nature of our objects in \autoref{sec:discussion}, where
  they lie in the centimeter luminosity vs.\ bolometric luminosity plot, how
  we can model their emission and how all the results affect the formation
  mechanisms of proto-\acp{BD}.
  Finally, \autoref{sec:conclusions} presents the conclusions of our work.

\section{Observations}
 \label{sec:obs}

\subsection{JVLA}

  We observed 11 proto-BD candidates in Taurus at 1.3 and 3.6-cm with the
  \acf{JVLA} of the \ac{NRAO}\footnote{The National Radio Astronomy Observatory
    is a facility of the National Science Foundation operated under
    cooperative agreement by Associated Universities, Inc.}.
  We selected the 11 sources in our initial sample that were associated with
  extended large scale emission in 250~$\mu$m maps of \emph{Herschel}, and
  thus could be associated with gas and dust emission of the Taurus cloud.
  In \autoref{fig:herschel}, we overlay the position of each \emph{Spitzer}
  source on the 250~$\mu$m maps of \emph{Herschel}, showing that all the
  \emph{Spitzer} sources fall in projection in regions of extended
  submillimeter emission, and several of them coincide with local emission
  enhancements at the position of the \emph{Spitzer} source.

  The 1.3-cm observations were carried out in 2013 June 10, while those at
  3.6-cm were performed in 2013 June 15.
  The correlator was set-up to use 8 GHz bandwidth per polarization for dual
  polarization mode at 1.3-cm, and 2 GHz bandwidth at 3.6-cm.
  Both sets of observations used 27 antennas in the VLA-C configuration. 
  We used J0336$+$3218 as gain calibrator and 3C147 as flux calibrator for
  both wavelengths.
  Each observing track was shared by all the sources in the sample. 
  J041757, our best proto-BD candidate, was observed at the beginning and at
  the end of each track for a total on-source time of $\sim8$ min at 1.3 cm
  and $\sim9$ min at 3.6 cm.
  The rest of the targets were observed for approximately $\sim5$ min
  on-source time at 1.3-cm and $\sim 4.5$ min at 3.6 cm.
  Pointing observations for the 1.3 cm data were done at the beginning and in
  the middle of the track, using the X band, on J0336$+$3218.
  The flux calibration observations on 3C147 at 1.3 cm were performed at the
  end of the track, using J0541$+$5312 as pointing source.
  The total observing time was 1h43min at 1.3~cm and 1h12min at 3.6~cm.

  Calibration and data reduction were performed using the \ac{CASA} package
  \citep{casaref} from the raw visibility data downloaded from the VLA archive
  as \ac{CASA} measurement sets and following \ac{NRAO} guidelines for
  calibration and imaging.
  We produced images using the common $uv$-range at 1.3 and 3.6~cm to sample
  comparable spatial scales at both wavelengths, and additionally used
  different weightings (Briggs's robust parameter ranging from 2, natural
  weighting to $-2$, uniform weighting), so that the final beams at 1.3 and
  3.6~cm are comparable.
  The average rms values are $\sim16~\mu$\jpb\ for 1.3 cm and
  $\sim30~\mu$\jpb\ for 3.6 cm.
  The average synthesized beam at 1.3~cm is $\sim 1.8''\times1.6''$.
  At 3.6~cm, the average beams are $\sim2.2''\times1.8''$ for uniform
  weighting and $\sim3.1''\times2.5''$ for natural weighting.

 \begin{table}
   \centering
   \caption{List of \emph{Herschel} observations used in this study} 

   \addtolength{\tabcolsep}{1pt}
   \begin{tabular}{lll}
     \linss
     \multicolumn{3}{c}{OBSIDs}  \\
     \linss
     1342190616  &  1342202090  &  1342202254  \\  
     1342202256  &  1342216549  &  1342216550  \\
     1342241898  &  1342241899  &  1342242047  \\
     \linss
   \end{tabular}
 \label{tab:herschel_obs}
 \end{table}

\subsection{Herschel Space Observatory}
 \label{sec:herschel_data}

  The Taurus molecular clouds were observed by the Herschel Space Observatory
  as part of the Gould Belt Survey \citep{Andre10}.
  A first set of observations was obtained in parallel mode using the PACS (at
  70 $\mu$m and 160~$\mu$m) and SPIRE (250~$\mu$m, 350~$\mu$m, and 500~$\mu$m)
  instruments simultaneously.
  The complete list of observations used in this study is reported in
  \autoref{tab:herschel_obs}. 
  More details about the observational strategy and depth of the maps can be
  found in \citet{Andre10}.
  The data were pre-processed using the Herschel Interactive Processing
  Environment \citep[HIPE,][]{Ott10} version 12, and the calibration files
  version 65 for PACS and version 22.0 for SPIRE.
  The final maps were subsequently produced using Scanamorphos version 24.0
  \citep{Roussel13}, using its galactic option, as recommended to preserve
  large scale extended emission.

\section{Results}
 \label{sec:results}

\subsection{1.3- and 3.6-cm emission}

  We detected emission over 3$\sigma$ at both 1.3 and 3.6~cm in five sources
  of our sample (J041757, J041847, J041913, J041938, and J042123); only at
  1.3~cm in J041836; and only at 3.6 cm in J041726 and J041740.
  For the rest of the candidates and/or bands, we only find upper limits.

 \begin{figure}
   \centering
   \includegraphics[width=\columnwidth]{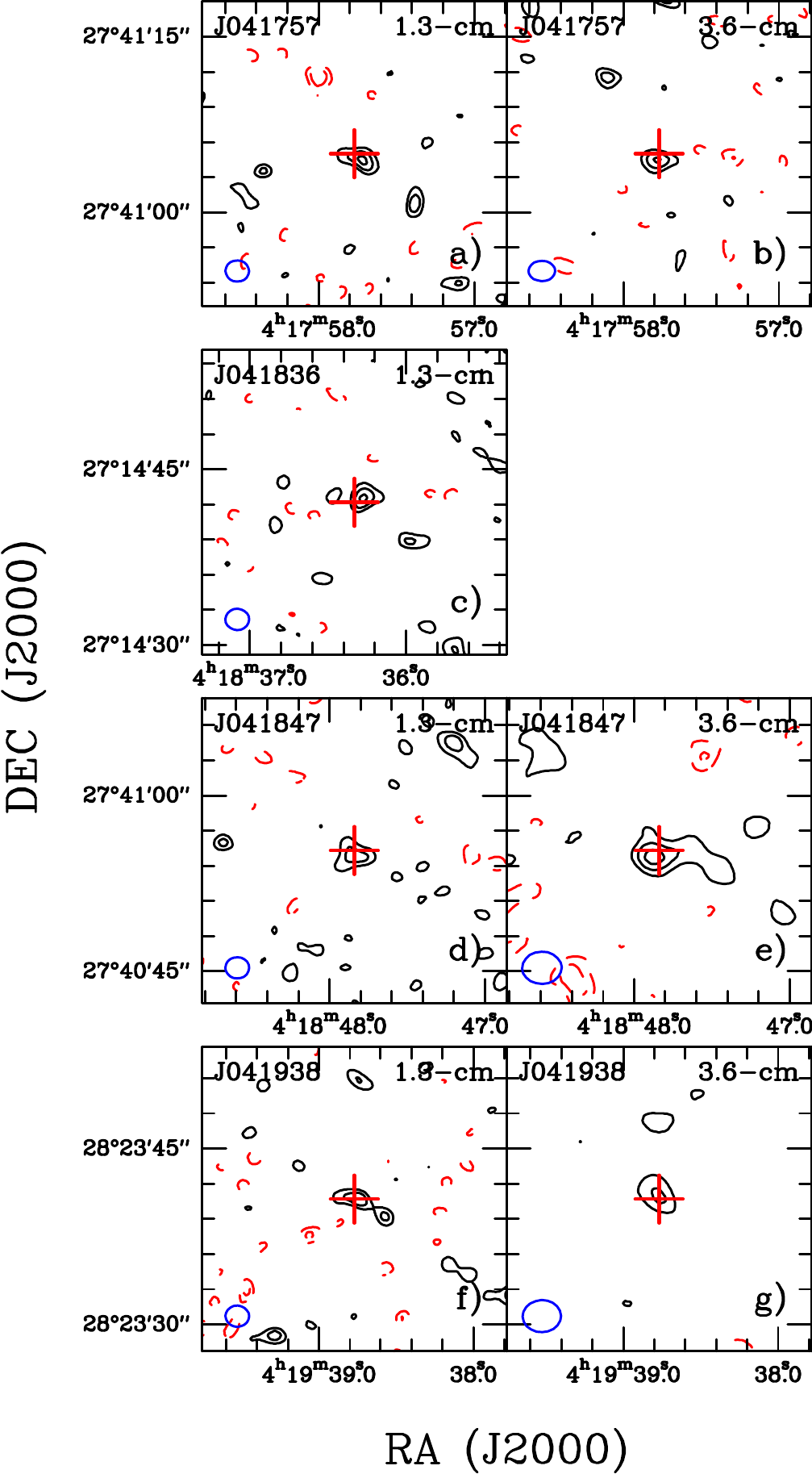}

   \caption{
     \ac{JVLA} maps of the 1.3- \textit{(left column)} and 3.6-cm
     \textit{(right column)} emission of the sources of our sample detected
     with partially resolved structures at 1.3 cm.
     Contours are -3,-2, 2, 3, \dots times the rms of the map, which from a)
     to g) are: 16, 31, 19, 20, 17, 20, and 25 $\mu$\jpb, respectively.  The
     red crosses mark the position of the Spitzer sources presented in
     \protect{\citet{Barrado09}}.  The blue ellipse at the lower left corner
     of each panel indicates the beam size.
   }

 \label{fig:map_extended}
 \end{figure}

 \begin{figure}
   \centering
   \includegraphics[width=\columnwidth]{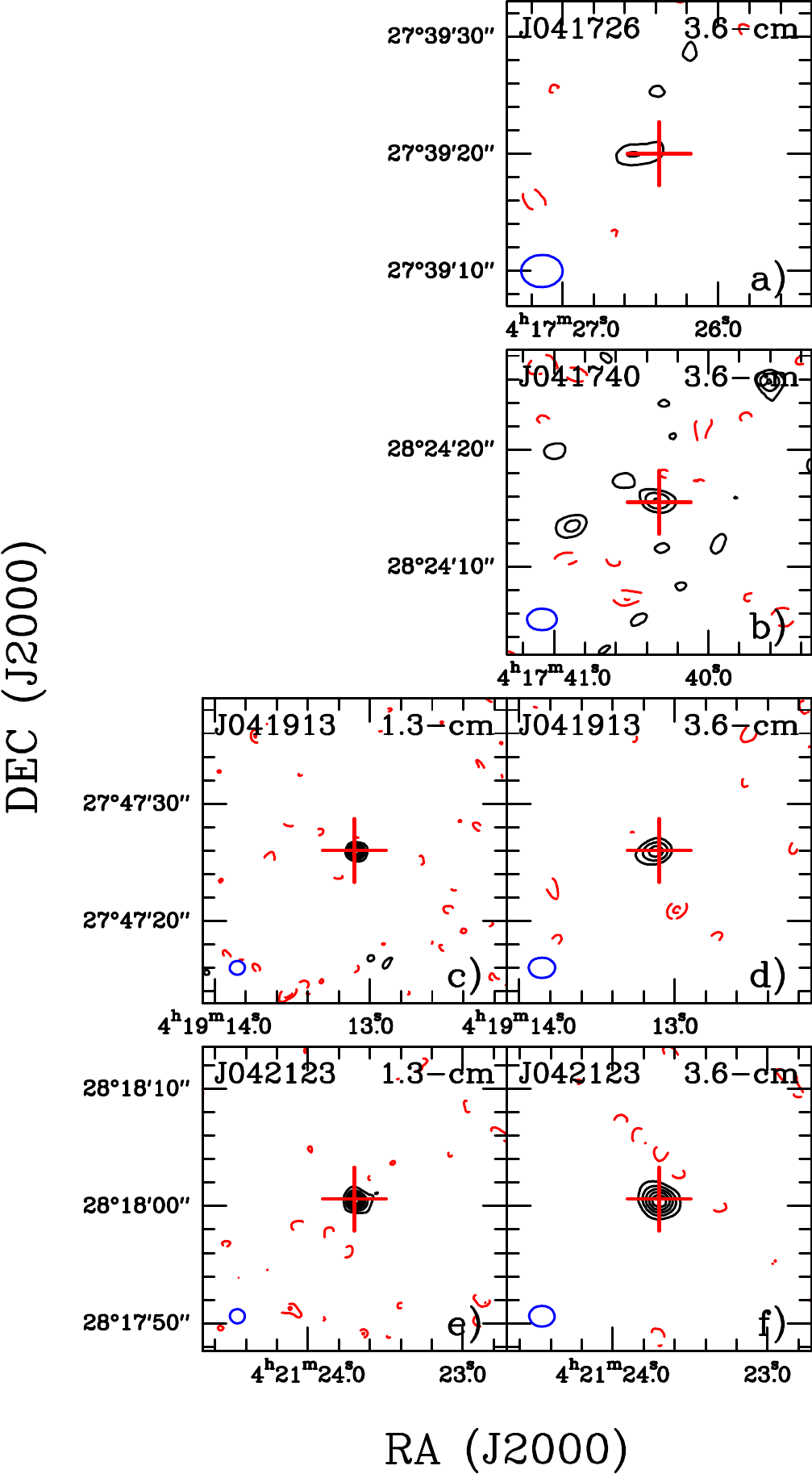}

   \caption{\ac{JVLA} maps of the 1.3- \textit{(left column)} and 3.6-cm
     \textit{(right column)} emission of the detected sources with point-like
     or undetected emission at 1.3 cm.
     Contours are: \textit{a)} $-3$,$-2$, 2, 3, 4; \textit{b)} $-3$,$-2$, 2,
     3, 4; \textit{c)} $-3$, $-2$, 3, 6, \dots; \textit{d)} $-3$, $-2$, 3, 4,
     5; \textit{e)} $-3$, $-2$, 4, 9, \dots; \textit{f)} $-3$, $-2$, 3, 6,
     \dots times the rms of the map, which from a) to f) are: 20, 25, 16, 50,
     16, and 51 $\mu$\jpb, respectively.
     The red crosses mark the position of the Spitzer sources presented in
     \protect{\citet{Barrado09}}.
     The blue ellipse at the lower left corner of each panel indicates the
     beam size.
   }

 \label{fig:map_pointsources}
 \end{figure}

  \autoref{fig:map_extended} presents the maps of the four sources in our
  sample that show slightly extended and faint ($\sim0.1$ m\jpb) emission at
  1.3~cm: J041757, J041836, J041847, and J041938.
  For these sources, we fitted a 2D Gaussian function, and report the position
  and deconvolved sizes in \autoref{tab:parameters}.
  The typical deconvolved sizes are $\sim 3''\times2''$, and the resulting
  positions of the sources match very well the position of the \emph{Spitzer}
  sources, within the uncertainties.
  In addition, these four slightly extended 1.3~cm sources show weak and
  almost unresolved emission at 3.6~cm (except for J041836, not detected at
  3.6~cm).

  \autoref{fig:map_pointsources} shows the maps of the sources with unresolved
  emission.
  J041740 is detected at 4$\sigma$ at 3.6 cm only, and is almost unresolved,
  while J041726 is just barely detected at 3.2$\sigma$ at 3.6 cm.
  The sources J041913 and J042123 are clearly detected at both 1.3 and 3.6~cm,
  with intensity peaks between 0.3 and 0.9~m\jpb, and signal-to-noise ratios
  $\sim8$--25.
  The emission from these last two unresolved sources is significantly more
  intense than the emission of the rest of the detected sources of our sample.
  The positions of these two sources also agree very well with the positions
  of the \emph{Spitzer} sources.

  Columns 9 and 10 of \autoref{tab:parameters} list the peak intensities and
  flux densities calculated for all the detected sources (or the corresponding
  upper limits) measured inside the 1$\sigma$ contour.
  The flux densities for the sources with partially resolved emission are
  between 0.09 and 0.15 mJy, while the two bright unresolved sources have
  larger flux densities by factors of 2 to 8.

  Column 11 of \autoref{tab:parameters} shows the calculated spectral indices
  from the flux densities measured in the detected sources of our sample.
  We calculate the spectral index, $\alpha$ as \citep{Kraus1986}
  \begin{equation}
    \alpha = \frac{\mathrm{ln}(S_{\nu_1}/S_{\nu_2})}
           {\mathrm{ln}(\nu_1/\nu_2)} 
  \end{equation}
  \noindent where $S_{\nu}$ is the measured flux density at a given frequency
  $\nu$.
  For the sources with detection at only one frequency, we used an upper limit
  of 3$\sigma$ for the flux density.
  The resulting spectral indices show that the four sources with faint and
  partially resolved 1.3-cm emission have spectral indices compatible with
  being $>-0.1$.
  The uncertainties in the determination of the spectral indices are
  relatively large, given the low S/N of most of the detections, which will
  only be improved with new and more sensitive observations.
  At the same time, the ratios between the flux densities at 1.3 and 3.6 and,
  in the case of J041757, an independent measure of the spectral index
  confirming the result of \citet{Palau12}, indicate that it is unlikely that
  the spectral indices of these sources are significantly $<-0.1$.
  On the other hand, the spectral indices for the two point-like sources
  J041913 and J042123, and for J041740, are clearly negative, and remain
  $<-0.1$ even taking into account the associated uncertainties.
  J041726, which is barely detected at 3.6-cm, shows a positive upper limit
  for the spectral index, but in this case we only have an upper limit for the
  emission at 1.3 cm.

\subsection{Spectral energy distribution of the sources}
 \label{sec:sed}

\begin{figure*}
   \centering
    \includegraphics[width=\textwidth]{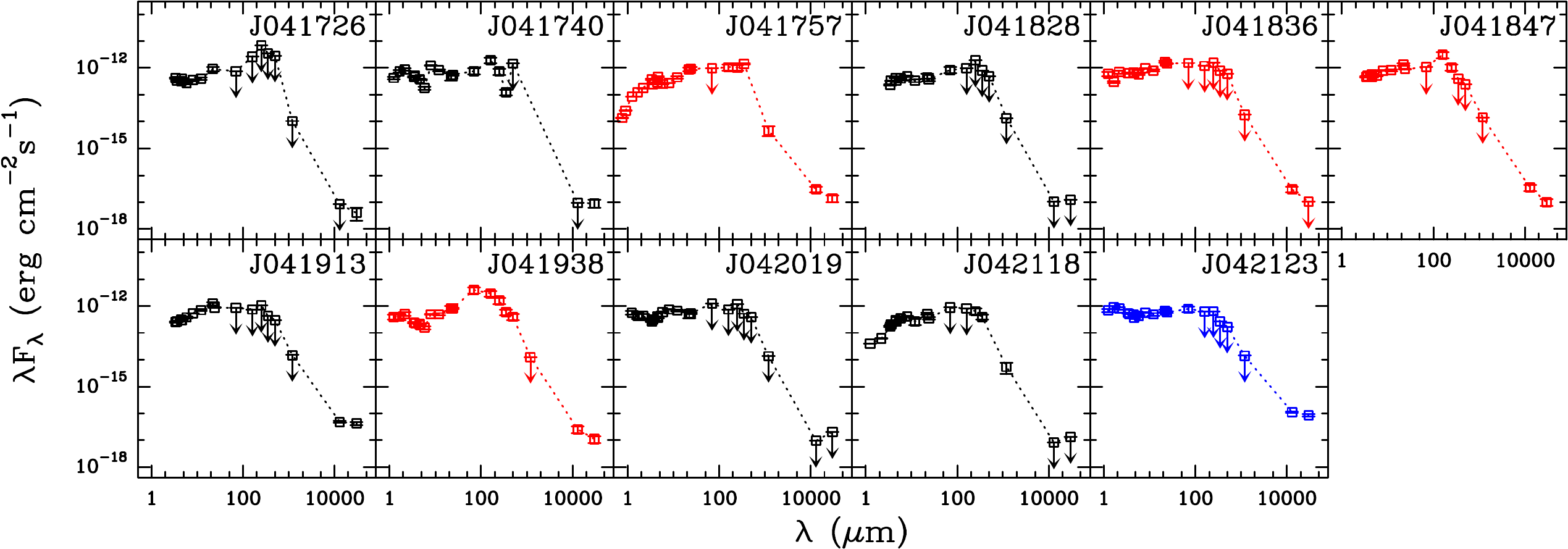}

    \caption{\acp{SED} for the 11 proto-\ac{BD} candidates studied in this
      work.  
      For most objects, \emph{Herschel} PACS detects a source. 
      The objects with \acp{SED} in red are those for which a thermal radio jet
      has been detected.
      The object with the \ac{SED} in blue corresponds to an identified
      radiogalaxy, through 21-cm continuum NVSS archival data (see
      \autoref{sec:nature}).
     }

 \label{fig:seds}
 \end{figure*}

  In \autoref{fig:seds}, we present the \acp{SED} of the sources of our
  sample, built using UKIDSS, 2MASS \citep{Skrutskie06}, and WISE databases,
  and measuring the fluxes in \emph{Spitzer} (IRAC and MIPS), and
  \emph{Herschel} (PACS and SPIRE) archive images (see
  \autoref{sec:herschel_data}).
  We additionally included the measurements at 350~$\mu$m reported in
  \citet{Palau12}, and the data presented in this work using the \ac{JVLA} at
  1.3 and 3.6~cm.
  All these data are listed in \autoref{apx:sed_data}.
  The \emph{Herschel} fluxes were obtained via aperture photometry.
  The values used for the apertures, inner and outer annulus radii of the
  background ring are (in this order): 10$''$, 10$''$, 20$''$ for all PACS
  bands, 20$''$, 20$''$, and 25$''$ for SPIRE 250 and 350 $\mu$m, and 30$''$,
  30$''$, 50$''$ for SPIRE 500 $\mu$m.
  The corresponding aperture correction factors were applied.
  When no detection is present in the \emph{Herschel} maps (after visual
  inspection), upper limits were computed as the standard deviation of several
  aperture photometry measurements performed around the source coordinates.
  There are no detections in any \emph{Herschel} band for J041726, J041836,
  J041913, and J042019.

 \begin{table*}
   \caption{Parameters of the sources observed with the \ac{JVLA}}

   \centering
   \addtolength{\tabcolsep}{-.35pt}

   \begin{tabular}{@{}lccccccccccc@{}}
     \linss
     & & & & 
     \multicolumn{2}{c}{Deconvolved\tnmk{a}} 
     & & & & &\\
     Source  &
     Wavelength  &
     \multicolumn{2}{c}{Position\tnmk{a}}  &
     ang.\ size &
     P.A.   &
     $\Tbol$\tnmk{b}  &
     $\Lbol$\tnmk{b}   &
     $I_{\nu}$  &
     $S_{\nu}$\tnmk{c}  &
     $\alpha$  &
     Type of \\
     & 
     (cm)  &
     $\alpha$ (J2000)  &
     $\delta$ (J2000)  &
     (arcsec)  &
     (deg.)  &
     (K)  &
     (\lo)  &
     (m\jpb)  &
     (mJy)  &
     &
     source \\
     \linss
     J041726  &  1.3  &
        & & & &  $>197$  &  $<0.0015$  &  $<0.036$\tnmk{d}  &
       $<0.04$\tnmk{e} &  $<0.16\pm$$0.74$\tnmk{e}  &  ? \\ 
     &  3.6  &
       4:17:26.50  &  27:39:20.0  &  \multicolumn{2}{c}{point source}  &  &  &
       0.05$\pm$0.02  &  0.04$\pm$0.02  & \\
     J041740  &  1.3  &
       & & & &  $\phantom{>}$389  &  $\phantom{<}$0.0032  &  $<0.040$\tnmk{d}
       &  $<0.04$\tnmk{e}  &  $<-$0.94$\pm$0.69 \tnmk{e}  &  ?\\  
     &  3.6  &
       4:17:40.34  &  28:24:15.7  &  $2.86\times0.07$  &  ~~57.3  &   &  &
       0.10$\pm$0.03  &  0.09$\pm$0.03  & \\ 
     J041757  &  1.3  &
       4:17:57.73  &  27:41:04.5  &  $3.25\times0.98$  &  ~~65.6 &
       $\phantom{>}$226  &  $\phantom{<}$0.0036  &  0.07$\pm$0.02  &
       0.13$\pm$0.03  &  $-$0.06$\pm$0.45  &  radio jet \\
     &  3.6  &
       4:17:57.78  &  27:41:04.4  &  $2.03\times0.73$  &  $-88.9$  &  &  &
       0.13$\pm$0.03  &  0.14$\pm$0.04  &  \\
     J041828  &  1.3  &
       &  &  &  &  $\phantom{>}$249  &  $\phantom{<}$0.0011  &
       $<0.045$\tnmk{d}  && \\
     &  3.6  &
       &  &  &  &  &   & $<0.123$\tnmk{d} & & &  ?\\
     J041836  &  1.3  &
       4:18:36.28  &  27:14:42.6  &  $2.05\times1.36$  &  $-68.1$  &  $>377$
       &  $<0.0033$  &  0.08$\pm$0.02  &  0.13$\pm$0.03  &
       $\phantom{;}>$0.02$\pm$0.47\tnmk{f}~~  &  radio jet\\ 
     &  3.6  &
       &  &  &  &  &   & $<0.105$\tnmk{d} &  $<0.12$\tnmk{f} & \\
     J041847  &  1.3  &
       4:18:47.84  &  27:40:54.9  &  $3.28\times2.12$  &  ~~63.8  &
       $\phantom{>}$126  &  $\phantom{<}$0.0041  &  0.08$\pm$0.02  &
       0.15$\pm$0.04  &  ~~0.43$\pm$0.42  &  radio jet \\ 
     &  3.6  & 
       4:18:47.86  &  27:40:54.8  &  $3.79\times1.38$  &  $-81.4$  &  &  &
       0.08$\pm$0.02  &  0.10$\pm$0.03  & \\
     J041913  &  1.3  &
       4:19:13.09  &  27:47:25.9  &  \multicolumn{2}{c}{point source}  &
       $>201$  &  $<0.0018$ &  0.28$\pm$0.02  &  0.28$\pm$0.02  &
       $-$0.44$\pm$0.16  &  ? \\
     &  3.6  &
       4:19:13.13  &  27:47:25.9  &  \multicolumn{2}{c}{point source}  &  &  &
       0.42$\pm$0.05  &  0.42$\pm$0.05  & \\
     J041938  &  1.3  &
       4:19:38.77  & 28:23:40.7  &  $4.01\times0.58$  &  ~~68.4 &
       $\phantom{>}$147  &  $\phantom{<}$0.0062  &  0.08$\pm$0.02  &
       0.11$\pm$0.03  &  ~~0.09$\pm$0.50  &  radio jet \\
     &  3.6  &
       4:19:38.78  &  28:23:41.0  &  $4.50\times0.23$ &  ~~36.9  &  &  &
       0.08$\pm$0.03  &  0.10$\pm$0.03  & \\
     J042019  &  1.3  &
       & & & &  $>487$  &  $<0.0018$  & $<0.041$\tnmk{d}  &  & &  ? \\
     &  3.6  &
       &  &  &  &  &  &  $<0.205$\tnmk{d}  &   & \\
     J042118  &  1.3  &
       & & & &  $\phantom{>}$166  &  $\phantom{<}$0.0020  &  $<0.036$\tnmk{d}
       & & & ? \\ 
     &  3.6  &
       &  &  &  &  &  &  $<0.132$\tnmk{d}  && \\
     J042123  &  1.3  &
       4:21:23.69 &  28:18:00.4  &  \multicolumn{2}{c}{point source}  &
       $\phantom{>}$646  &  ---\tnmk{g}  &  0.53$\pm$0.02  &  0.53$\pm$0.02  &
       $-$0.64$\pm$0.09  & radiogalaxy \\
     &  3.6  &
       4:21:23.70  &  28:18:00.4  &  \multicolumn{2}{c}{point source}  &  &
       &  0.94$\pm$0.05  & 0.94$\pm$0.07  &  \\ 
     \linss
   \end{tabular}

   \begin{minipage}{\textwidth}

     \tntxt{a}{Obtained from fitting a 2D Gaussian function to the emission.}

     \tntxt{b}{\hang $\Tbol$ is calculated from the \ac{SED} following
       \citep{Chen95} and $\Lbol$ is calculated integrating the \ac{SED}, and
       assuming a distance $D=140$ pc.
       For J041726, J041836, J04193 and J042019, which are not detected in any
       \emph{Herschel} band, we estimated $\Tbol$ and $\Lbol$ using the upper
       limit of PACS at 70 $\mu$m, and considered the resulting $\Tbol$ as a
       lower limit and the resulting $\Lbol$ as an upper limit.}

     \tntxt{c}{Flux densities were measured inside the 1$\sigma$ contour of
       the emission.}

     \tntxt{d}{Upper limit calculated as 3$\sigma$, where $\sigma$ is rms of
       the map.}

     \tntxt{e}{Calculated using an upper limit for the 1.3-cm flux density,
       S$_{\nu}$, $S_\mathrm{lim} = 3\sigma$, where $\sigma$ is the rms of the
       map \citep{Beltran01}.}

     \tntxt{f}{\hang Calculated using an upper limit for the 3.6-cm flux
       density, S$_{\nu}$, $S_\mathrm{lim} = 3\sigma A^{0.7}$, where $\sigma$
       is the rms of the map, and $A$ is the source area in beam units
       estimated from the 1.3-cm observations \citep{Beltran01}.}

     \tntxt{g}{Given the classification of this object as a radiogalaxy, the
       $\Lbol$ calculated from the \ac{SED} is meaningless.}

   \end{minipage}

 \label{tab:parameters}
 \end{table*}

  \autoref{fig:seds} shows that the \acp{SED} are in general flat in the
  range 2--100~$\mu$m, or peaking around 100~$\mu$m in some cases, such as
  J041757, J041847, J041938 and J042118.
  Interestingly, out of the 4 sources with \acp{SED} peaking around
  100~$\mu$m, 3 have flat or positive spectral indices in the centimeter
  range.
  In order to estimate this in a more quantitative way, we calculated the
  bolometric temperatures, $\Tbol$ (see \autoref{tab:parameters}), following
  \citet{Chen95}, and find them to range from 126~K to 650~K for all the
  sample.
  Thus, most of the sources of our sample present \acp{SED} comparable to
  Class I young stellar objects.
  Two of the sources where we detect thermal radio jets, J041847 and J041938,
  have the lowest values of $\Tbol$, with $\Tbol<150$~K, which also suggests
  that they are probably Class~0/I sources.
  J041757 is also probably a young Class~I object given the derived
  $\Tbol=226$~K.
  We also calculated the bolometric luminosities, $\Lbol$ (see
  \autoref{tab:parameters}), and find values ranging from 0.001 to 0.006~\lo,
  well within the proto-\ac{BD} regime (according, \eg\ to the evolutionary
  models of \citealt{Baraffe02}).

\section{Discussion}
 \label{sec:discussion}

\subsection{Nature of the objects}
 \label{sec:nature}

  We can divide the detected sources in two distinct classes according to the
  spectral index calculated from the 1.3- and 3.6-cm continuum emission.
  The sources that show intense and unresolved emission, or no emission at 1.3
  cm, J041740, J041913 and J042123, have clearly negative spectral indices
  consistent with being originated by non-thermal synchrotron emission,
  $\alpha \sim -0.6$ \citep[\eg][]{BiegingCohen89,Girart02,Dzib13}.
  This emission is seen in T~Tauri stars and radiogalaxies.
  In order to account for an extragalactic origin for our sources, we
  searched the VLA archive and the NRAO VLA Sky Survey (NVSS) \citep{NVSS98}
  to check if there were any counterparts at longer wavelengths.
  We only found detections at several epochs of 21-cm radio continuum emission
  for J042123, which show a central object with two radio lobes on opposite
  sides of it, clearly tracing a radiogalaxy.
  We also searched the NASA Extragalactic Database for possible counterparts
  for our sources and only found J042123 to have an extragalactic counterpart,
  given the typical positional uncertainties in WISE and GALEX ($<1''$) and
  the positional uncertainties of our sources (determined from Spitzer/IRAC
  and the VLA, $<1''$).
  Thus, we classify J042123 as a radiogalaxy and do not consider it a proto-BD
  candidate any more.
  The classification of J041740 and J041913 remains unsettled due to lack of
  information.

  On the other hand, the spectral indices for the four sources detected at 1.3
  cm, with partially resolved emission, are compatible with flat or positive
  spectral indices, $>-0.1$ (already reported by \citealt{Palau12} for the
  case of J041757).
  These spectral indices are expected for optically thin ($\alpha\approx-0.1$)
  to partially optically thick thermal free-free emission.
  The origin of the thermal free-free emission in low-mass young stellar
  objects is usually related to shocks generated in an outflow
  \citep[\eg][]{Curiel89,Beltran01,GonzalezCanto02,Lynch13}, and this kind of
  sources are designated as `thermal radio jets'.
  The 1.3-cm emission from these four objects also shows a marginal degree of
  elongation, which supports the idea that the emission is originated in a jet
  emanating from the substellar object.
  Thermal radio jets from young stellar objects are more easily seen at the
  most embedded Class 0/I phases, and three of the four objects where we
  find thermal radio jets present $\Tbol$ in the range 130--230~K, consistent
  with Class~I phase $\Tbol$ near the border with the Class~0 phase.
  Thus, the detection of partially elongated thermal emission from several
  objects in our sample makes these four objects excellent candidates to Class
  I proto-\acp{BD}.

\subsection{Centimeter vs.\ bolometric luminosity}

\begin{figure}
   \centering
   \includegraphics[angle=-90,width=\columnwidth]{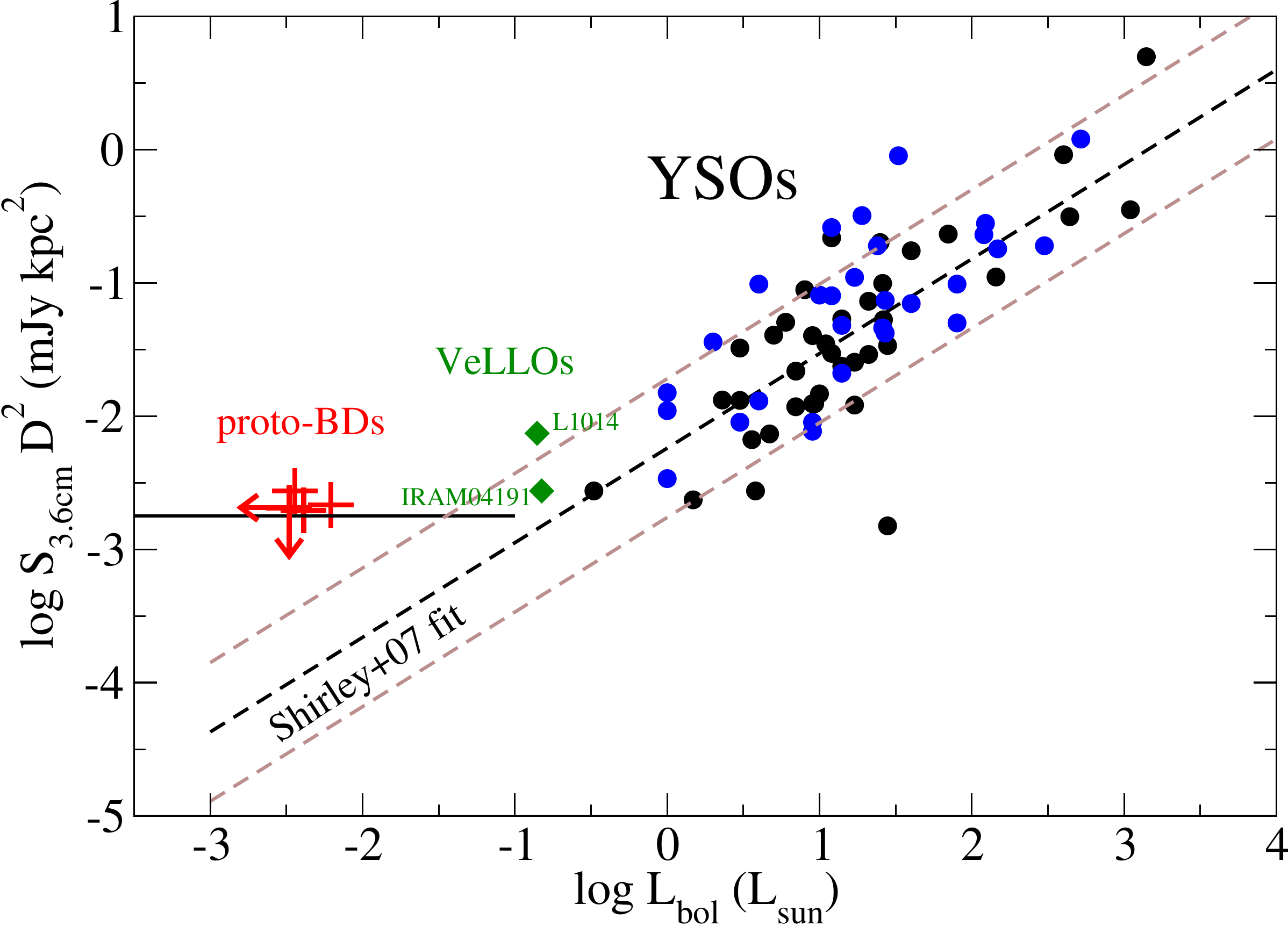}

   \caption{
     Centimeter luminosity at 3.6 cm vs bolometric luminosity.
     Blue and black dots correspond to the data compiled by
     \protect{\citet{Anglada95}} and \protect{\citet{Furuya03}}, respectively,
     showing the relation for \acp{YSO}.
     Red plus signs correspond to the proto-\ac{BD} candidates driving
     radio jets presented in this work.
     \acp{VeLLO} with detected 3.6~cm emission
     \protect{\citep{Andre99,Shirley07}} are shown as green squares.
     The horizontal black line indicates the typical detection limit from our
     3.6-cm observations, calculated as the 3$\sigma$ value of the typical
     rms of our 3.6-cm maps, $\sim30~\mu$\jpb.
     The black dashed-line is the fit performed by \protect{\citet{Shirley07}}
     to the \acp{YSO}.
     The two brown dashed lines indicate the standard deviation of the fit
     obtained by \protect{\citet{Shirley07}}, calculated as
      $(\sum\limits_{i}(L_\mathrm{cm}^{i}(obs)-L_\mathrm{cm}^{i}(fit))^2/N)^{1/2}$.
     }

 \label{fig:LcmLbol}
 \end{figure}

  \autoref{fig:LcmLbol} shows the centimeter luminosity at 3.6 cm measured in
  our \ac{JVLA} observations versus the bolometric luminosity calculated from
  the \acp{SED} of the objects of our sample associated with thermal radio
  jets and assuming that they belong to Taurus.
  We compare the results for our sample with the same variables measured for a
  sample of \acp{YSO} from \citet{Anglada95} and \citet{Shirley07}, where
  we also include the values for known \acp{VeLLO} with detected 3.6~cm
  emission \citep{Andre99,Shirley07}.

  In general terms, the results for our sample of proto-\ac{BD} candidates
  follow the trend expected for \acp{YSO}, extending to $\Lbol$ about one
  order of magnitude smaller.
  The relationship between the centimeter continuum luminosity and the
  bolometric luminosity of low-mass protostellar objects is interpreted to
  arise from the intrinsic relation between the stellar wind properties and
  the stellar mass. 
  Since the stellar wind shocks against the surrounding high density material,
  ionizing the gas, more luminous young stellar objects are expected to have
  higher centimeter luminosity, due to stronger stellar winds and probably
  denser surrounding gas \citep[see \eg][]{Curiel87,Curiel89}.
  However, the centimeter luminosities of the thermal radio jets of our sample
  (red crosses) present a systematic excess of about an order of magnitude
  from the relation found for \acp{YSO} (dashed line in
  \autoref{fig:LcmLbol}).
  This excess seems to be significant because the plotted bolometric
  luminosity is an upper limit to the true luminosity of the object (better
  approximated by the internal luminosity, \citealt{DiFrancesco07}).

  However, we must be careful with our interpretation. 
  The detected 3.6-cm luminosities of the thermal radio jets in our sample
  are very close to the detection limit we estimate from the rms of the
  undetected sources in our sample.
  With our data, we cannot discard the possibility that we are only seeing the
  upper tip of the more brilliant radio jets of a population of proto-\acp{BD},
  while we cannot detect the emission of weaker radio jets that would lie
  closer to the fit of \citet{Shirley07}.

  In order to account for this `extra' centimeter emission, we discarded
  several mechanisms discussed in the literature, such as supersonic accretion
  onto a protostellar disk \citep{NeufeldHollenbach96,Shirley07}, because they
  do not seem feasible at all for the masses and accretion rates of the
  \ac{BD} regime.
  Similarly, it seems very unlikely that the four detected radio jets in our
  sample have a burst, at the same time, of a variable (non-thermal) component
  with respect to the quiescent (thermal) component as found in the
  \acp{VeLLO} L1014-IRS \citep{Shirley07} and IRAM04191 \citep*{Choi14}.

  One possibility left is that the thermal radio jet or wind emanating from a
  proto-\ac{BD} impacts against a medium with a higher density as compared to
  the case of low-mass star formation.
  This could be expected, as proto-\acp{BD} should not be highly efficient in
  creating cavities through the passage of their outflows.
  In \autoref{sec:models}, we use the same type of models used to reproduce
  the emission of thermal radio jets originating from low-mass \acp{YSO}
  and adapt them to the physical parameters of the \ac{BD} regime, in order to
  explain the flux densities we detected for our thermal radio jets.

\subsection{Models of emission}
 \label{sec:models}

  Here we explore theoretical models which have been used to explain the
  emission of thermal radio jets from low-mass \acp{YSO}.
  We use the models developed by \citet{Curiel87}, \citet{GonzalezCanto02} and
  \citet{Rodriguez12} in order to explain the observed radio-continuum
  emission (with thermal origin) from the sources of our sample.
  \citet{Curiel87} calculated for the first time the radio-continuum emission
  (of thermal origin) produced by a plane-parallel shock wave.
  \citet{GonzalezCanto02} and \citet{Rodriguez12} modeled the observed radio
  emission in low-mass stars as internal shocks, which are produced by
  intrinsic variability in the injection velocity, in a spherical wind and a
  bipolar outflow (with conical symmetry), respectively.
  We follow the approach of these authors to explore three possible geometries
  for the thermal emission detected in our sample of proto-\ac{BD} candidates,
  since our observations are not able to fully resolve the geometry of the
  emitting region or to measure the degree of collimation of the thermal radio
  jets.
  We first calculate the emission produced by a plane-parallel shock wave in
  \autoref{sec:model0}.
  We then apply the case of a stationary stellar wind with spherical symmetry
  in \autoref{sec:model1}.
  Given that the partially resolved 1.3-cm emission maps show marginal
  elongations along a preferred direction, we also present in
  \autoref{sec:model2} the results of a variable conical wind model,
  collimated in a similar way as the \acp{YSO} winds.
  The assumed distance for all the models is $D=140$~pc, the adopted distance
  to the Taurus cloud \citep{Loinard05}.

\subsubsection{Plane-parallel shock wave}
 \label{sec:model0}

 \begin{figure}
   \centering
   \includegraphics[width=6cm]{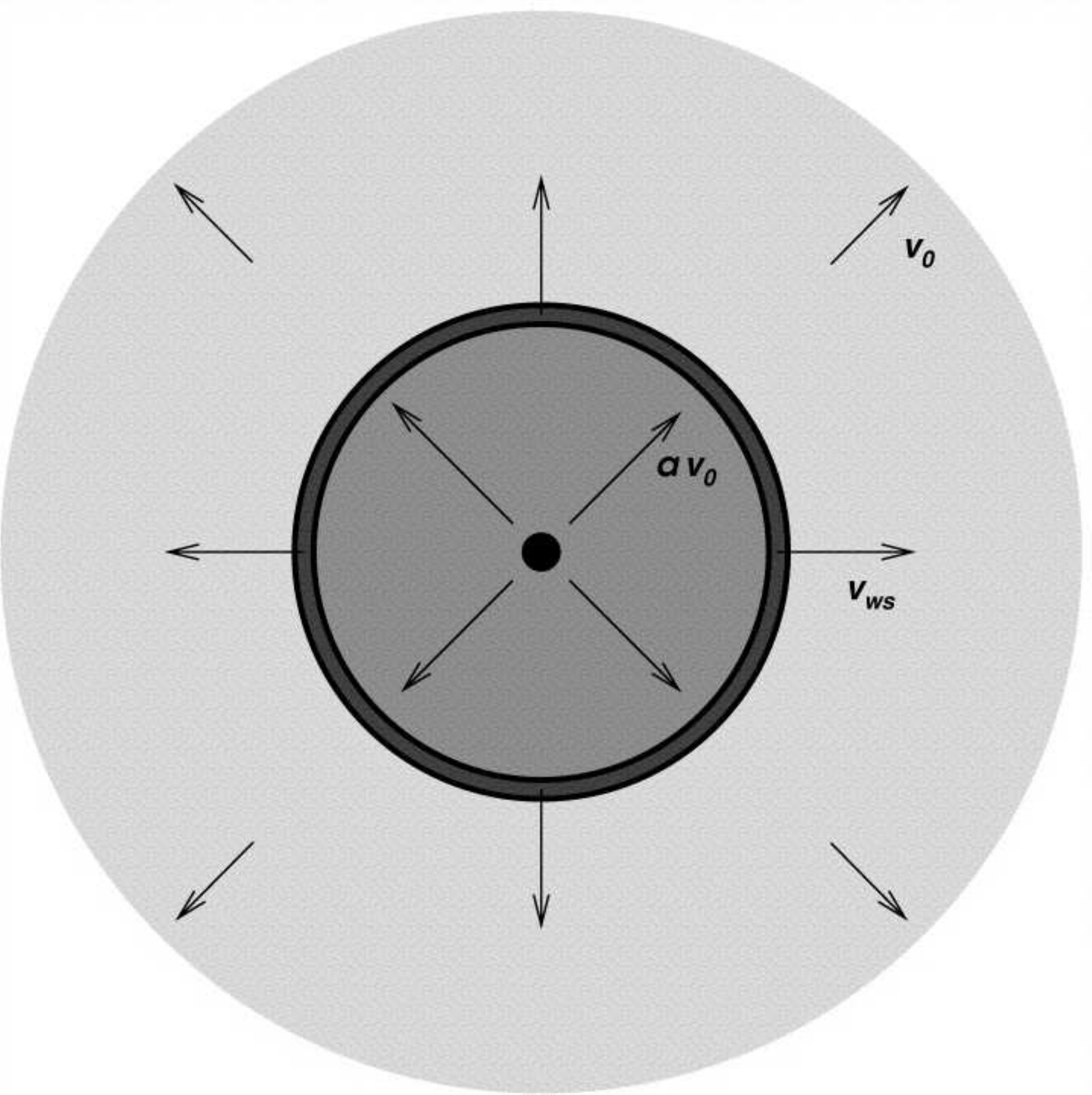}
   \includegraphics[width=\columnwidth]{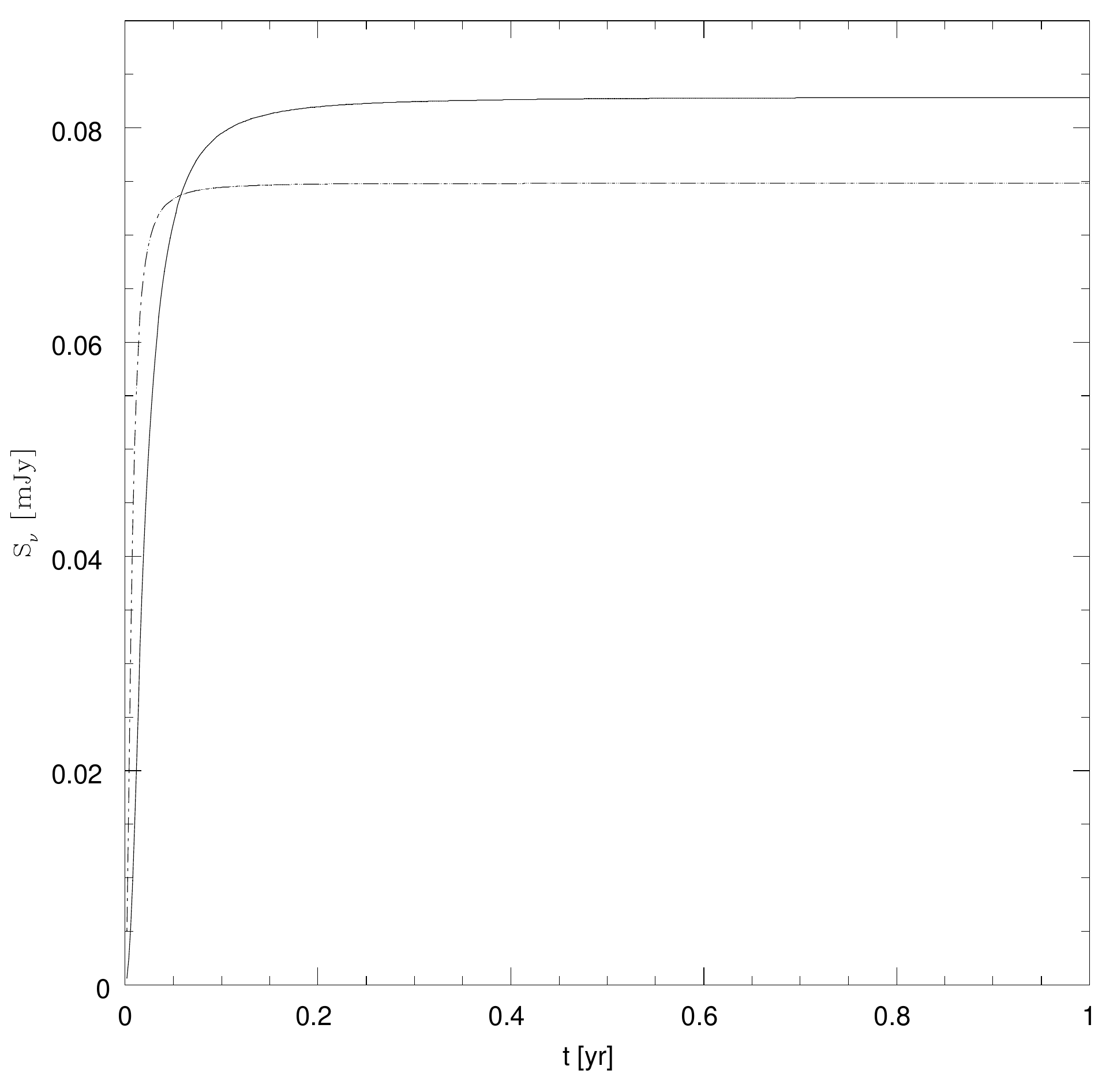}

   \caption{
     \emph{Top:} Schematic diagram showing a stellar wind with velocity $V_0$
     which suddenly changes its value to $a\,V_0$ ($a > 1$).
     The interaction between these outflows produces an internal working
     surface that propagates outwards with an intermediate speed $V_{WS}$.
     Figure taken from \protect{\citet{GonzalezCanto02}}.
     \emph{Bottom:} Predicted flux densities at $\lambda =$1.3 cm (dashed
     line) and $\lambda =$3.6 cm (solid line) from a spherical wind with an
     internal single working surface.
     We have adopted an initial ejection velocity $V_0 = 50$~\kms, which
     suffers a sudden increase to $a\,V_0 = 200$~\kms.
     The mass-loss rate $\dot m = 1.0 \times 10^{-8}$~\moyr\ remains constant.
   }

 \label{fig:model1} 
 \end{figure}

  \citet{Curiel87} developed an analytic model in order to calculate the
  free-free emission at radio frequencies produced in plane-parallel shock
  waves.
  These authors assumed that radiation is produced in the recombination zone,
  which is considered to be isothermal (T$\simeq$ 10$^4$ K).
  The flux density is calculated using a correlation between the radio
  intensity and H$_\beta$ intensity from the recombination zone.
  From an application of the model to an astrophysical source with a circular
  geometry, the flux density at radio frequencies in the optically thin regime
  ($\tau_{\nu}\ll$ 1, being $\tau$ the optical depth and $\nu$ the frequency)
  is given (in mJy) by,
 \begin{equation}
 \label{eq:flux}
   \begin{aligned}
     S_{\nu}= 1.84 \times 10^{-4} \, \biggl(\frac{n}{10\,\mbox{cm}^{-3}}\biggr)
     \biggl(\frac{V}{100\,\mbox{km s}^{-1}}\biggr) \\
     \times\,\biggl(\frac{\theta}{\mbox{arc sec}}\biggr)^2
     \biggl(\frac{\nu}{10\, \mbox{GHz}}\biggr)^{-0.1} 
   \end{aligned}
 \end{equation}
  \noindent where $n$ is the pre-shock density, $V$ is the shock velocity, and
  $\theta$ is the angular diameter of the source.
  For shock velocities $V\sim100$~\kms, the number of ionizing photons per
  atom produced by the shock wave is $\phi \sim 1$ \citep[see, for
    instance,][]{Kang1992}.
  When we applied the model of \citet{Curiel87} to our proto-\ac{BD}
  candidates, we chose a source with an angular diameter $\theta = 2''$
  (\autoref{tab:parameters}).
  The adopted wind parameters\footnote{Mass outflow rates measured in several
    Class~II \protect{\acp{BD} range from $0.5$--$40\times10^{-9}$
      \moyr\ \protect{\citep{PhanBao2008,PhanBao2011,PhanBao2014a,Whelan2014b};
        while the only mass outflow rate measured for a possible Class~I
        \protect{\ac{BD}} is $1\times10^{-9}$
        \moyr\ \protect{\citet{Riaz2015}}.
    Typical velocities measured for winds of embedded \protect{\acp{BD}} range
    from 20 to 100~\kms
    \protect{\citep{Joergens12,Whelan2009a,Whelan2009b}}.}} are $\dot m$ = 2
    $\times 10^{-9}$ \moyr\ and $V = 100$~\kms.}
  The pre-shock density is calculated from the continuity equation, given the
  position of the shock wave with respect to the central source.
  Substitution of these values into \autoref{eq:flux} gives the flux density
  $S_{\nu}=$ 0.1 mJy at a frequency $\nu =$ 8.33 GHz.

\subsubsection{Stationary stellar wind with spherical symmetry}
 \label{sec:model1}

  Let us now consider a spherical stationary stellar wind characterized by an
  ejection velocity $V_0$ and mass-loss rate $\dot m$.
  At $t= 0$, the wind velocity suddenly increases to the value $a\,V_0$ (with
  $a > 1$), while the mass-loss rate remains constant.
  This kind of variability was previously studied by
  \citet{GonzalezCanto02,GonzalezCanto08}.
  Such variation in the wind parameters instantaneously produces (at the base
  of the wind) a two-shock wave structure (called a working surface) that
  propagates outwards over time with a speed $V_{WS} = a^{1/2}\,V_0$. 
  Note that this value is intermediate between the slow and faster wind
  velocities (see top panel of \autoref{fig:model1}).

  If we assume that the working surface is thin enough to be described by a
  unique distance $r_{WS} = r_{WS} (t)$, it is possible to determine the total
  optical depth $\tau_{WS} = \tau_{IS} + \tau_{ES}$ of the shocked layer,
  where $\tau_{IS}$ and $\tau_{ES}$ are the optical depths of the internal and
  external shocks, respectively, that bound the working surface\footnote{The
    optical depth of the shocks at radio frequencies has been estimated as
    $\tau= \beta\, n_0 \,V_s^{\gamma}\,\nu^{-2.1}$, where $n_0$ is the
    pre-shock density, $V_s$ is the shock velocity, and $\beta$ and $\gamma$
    are constants that depend on the shock velocities
    \protect{\citep{GonzalezCanto02}}.}.
  Then, we estimate the intensity emerging from each direction and calculate
  the flux density by integrating the intensity over the solid angle. That is,
 \begin{equation}
   S_{\nu}= 2 \pi B_{\nu} {\biggl ( \frac{r_{WS}}{D}\biggr )}^2 \int_0^1
   \biggl( 1 - e^{-2 \tau_{\nu}(\mu)} \biggr) \mu d\mu 
 \end{equation}
  \noindent where $B_{\nu} (= 2 k T_e \nu^2/ c^2$) is the Planck function in
  the Rayleigh-Jeans approximation, and $\mu$ = cos\,$ \theta$, being $ \theta$
  the angle formed by each line of sight and the normal to the working surface.

  The bottom panel of \autoref{fig:model1} shows the results for a model
  for the radio continuum emission from a single working surface.
  The density fluxes were computed at frequencies $\nu=$8.33~GHz
  ($\lambda=$~3.6 cm) and $\nu=$~23.08 GHz ($\lambda$ = 1.3 cm).
  We chose representative parameters for \ac{BD} sources: 
  \emph{i)} a stellar wind with an initial ejection velocity, $V_0 = 50$~\kms,
  that suddenly increases to 200~\kms\ ($a= 4$); and \emph{ii)} a constant
  mass-loss rate $\dot m = 1.0 \times 10^{-8}$ M$_\odot$ yr$^{-1}$, an order
  of magnitude higher than the values derived from the measurements of
  \citet{Lee13} and \citet{Palau2014} for two proto-\ac{BD} candidates.
  Using these values, we obtain a working surface velocity $V_{WS} =$
  100~\kms, and consequently, shock velocities of $V_{is} =$ 100~\kms\ and
  $V_{es} =$ 50~\kms\ for the internal shock and the external shock,
  respectively.
  At the beginning, the working surface is optically thick at both
  wavelengths, and the flux grows as $t^2$ (since $r_{WS}\propto t$).
  When the shocked layer becomes optically thin (the transition time depends
  on the frequency as $\nu^{-1.05}$), the flux densities tend to constant
  values $S_\mathrm{1.3~cm} \simeq 0.075$ mJy and $S_\mathrm{3.6~cm} \simeq
  0.083$ mJy, while the spectral index $\alpha_{1.3-3.6} \simeq$ tends to
  $-0.1$.

\subsubsection{Variable conical outflow}
 \label{sec:model2}

\begin{figure}
   \centering
   \includegraphics[width=8cm]{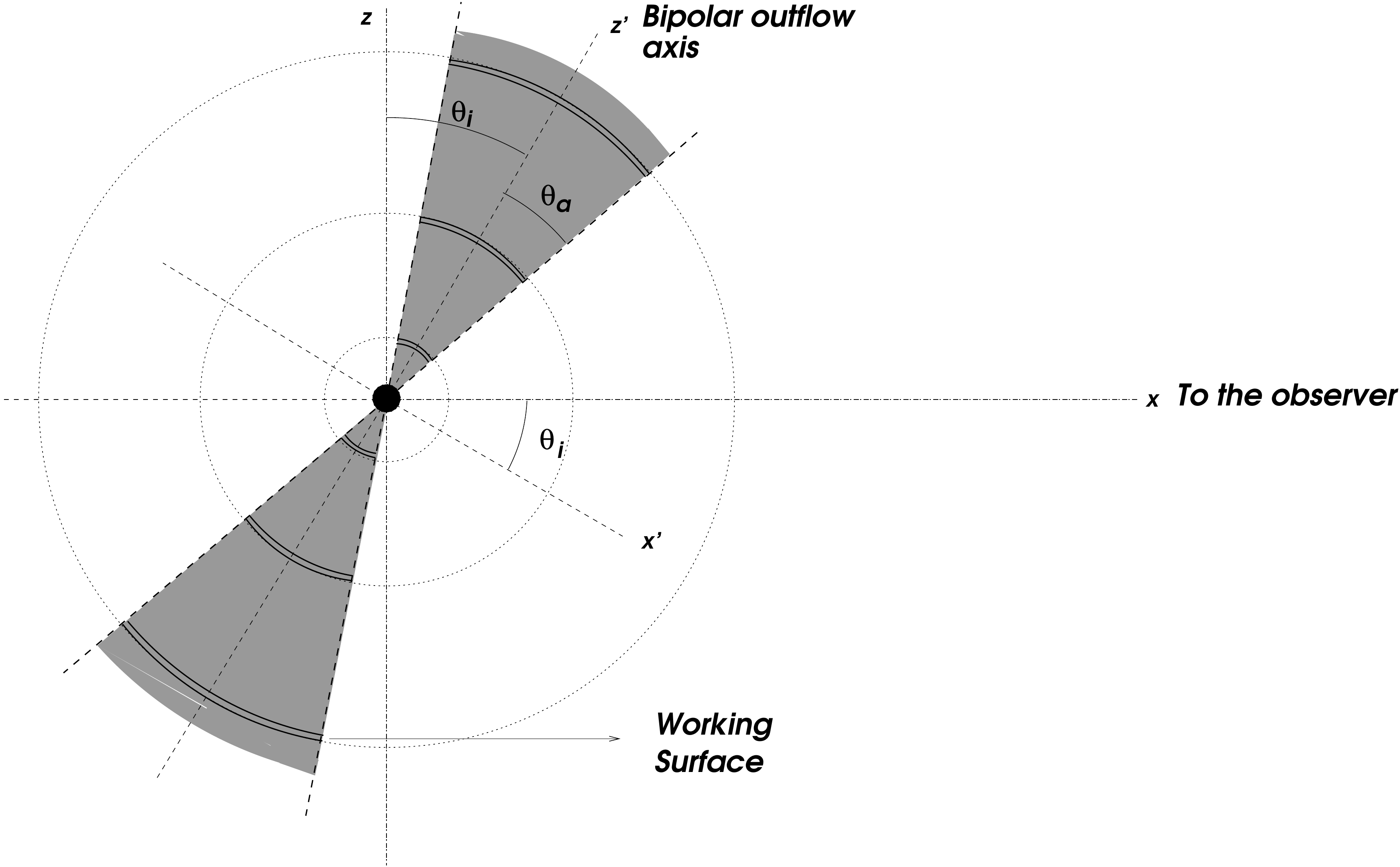}
   \includegraphics[width=\columnwidth]{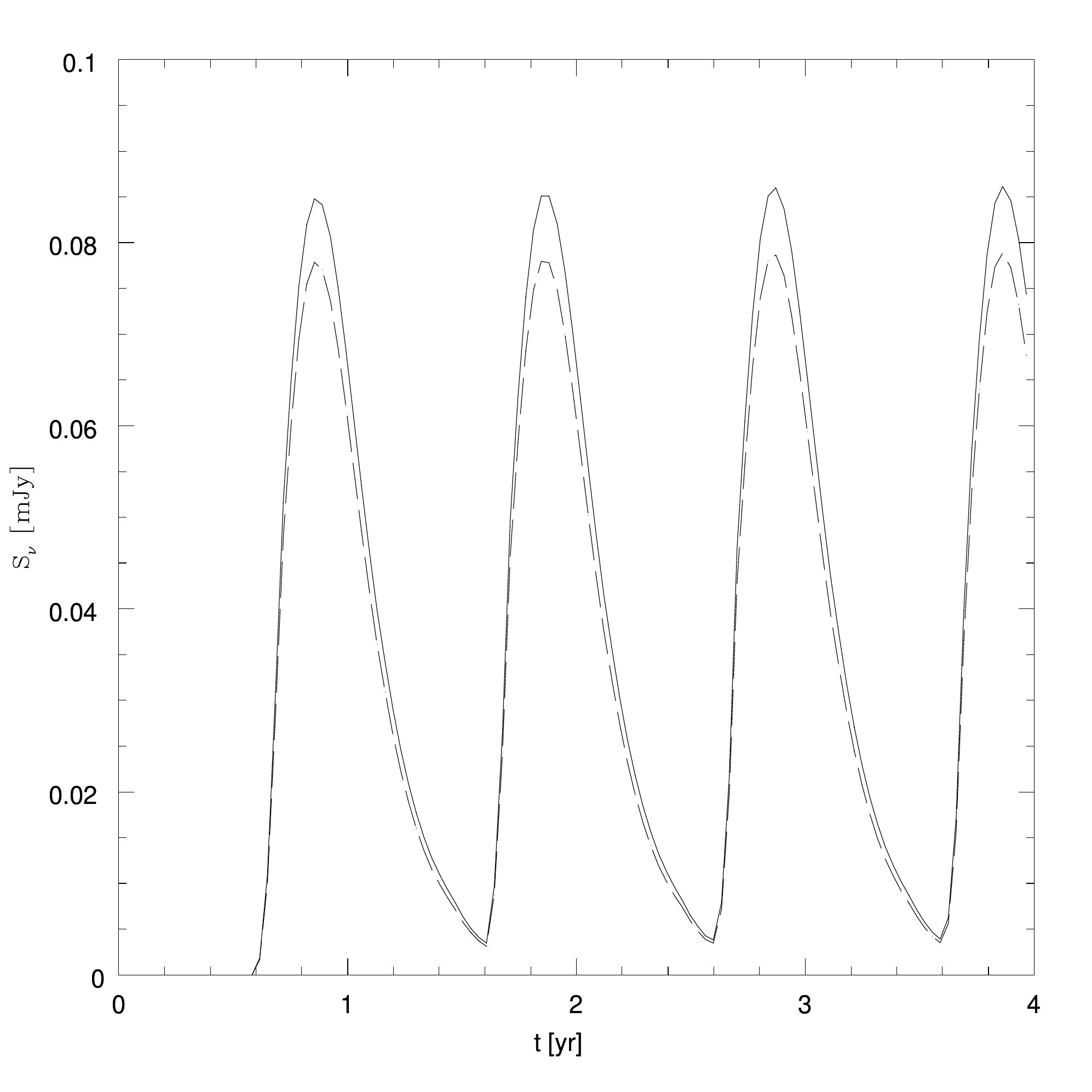}

   \caption{
     \emph{Top:} Schematic diagram showing a bipolar outflow with conical
     symmetry.
     The angles $\theta_a$ and $\theta_i$ are the opening angle of the cones
     and the inclination angle with the plane of the sky, respectively.
     Figure taken from \citet{Rodriguez12}.
     \emph{Bottom:} Predicted flux densities at $\lambda =$ 1.3 cm (dashed
     line) and $\lambda =$ 3.6 cm (solid line) for a bipolar outflow with a
     sinusoidal ejection velocity.
     We assumed the following wind parameters: mean velocity $V_w = 125$ km
     s$^{-1}$, amplitude $V_c = 75$~\kms, oscillation frequency $\omega$= 6.28
     yr$^{-1}$ (period $P=2\pi/\omega$= 1 yr), and constant mass-loss rate
     $\dot m = 5 \times 10^{-8}$ M$_\odot$ yr$^{-1}$. 
     The opening angle of the cones is $\theta_a= $ 45$^{\circ}$ with an
     inclination angle $\theta_i$= 45$^{\circ}$.
    }
 \label{fig:model2}
 \end{figure}

  We also computed the free-free emission from a bipolar outflow with conical
  symmetry.
  We assumed that the opening angle of the cones is $\theta_a$, and the
  inclination angle between the jet axis and the plane of the sky is
  $\theta_i$.
  We show a schematic diagram of the geometrical model in the top panel of
  \autoref{fig:model2}.

  For this model, we assumed a periodic variation in the injection velocity
  $V_{jet}$ and constant mass-loss rate $\dot m_{jet}$.
  In this scenario, the radiation is produced by internal working surfaces
  which move outwards over time from a central source.
  In particular, we consider a stellar conical outflow expelled with a
  sinusoidal variation in the injection velocity, $V_{jet}= V_w -V_c\,
  \mbox{sin} (\omega \tau)$, where $V_w$ is the mean velocity of the outflow,
  $V_c$ is the amplitude of the velocity variation, $w$ is the frequency, and
  $\tau$ is the injection time.

  In order to obtain the flux density from the bipolar outflow,
  \citet{Gonzalez02} and \citet{Rodriguez12} found the conditions that
  indicate whether or not a working surface is intersected by a given line of
  sight.
  For this goal, this model described the internal working surfaces as polar
  caps (portions of spheres) whose physical sizes depend on the opening angle
  $\theta_a$ and their position from the central source.
  First, the total optical depth along each line of sight is estimated adding
  the optical depths of the working surfaces intersected by each line of
  sight.
  Then, the intensity emerging from each direction is calculated, and finally,
  the total flux emitted by the system is computed by integrating this
  intensity over the solid angle.

  We present the predictions of our model in the bottom panel of
  \autoref{fig:model2}.
  We show a numerical example for the radio-continuum flux at $\lambda = 1.3$
  cm and $\lambda = 3.6$ cm.
  The opening angle of the cones is $\theta_a= $ 45$^{\circ}$ and the
  inclination angle is $\theta_i$= 45$^{\circ}$.
  We assumed a mean velocity $V_w = 125$~\kms, an amplitude $V_c = 75$~\kms,
  and oscillation frequency $\omega$= 6.28 yr$^{-1}$ (that is, a period
  $P=2\pi/\omega$= 1 yr) and a constant mass-loss rate $\dot m = 5 \times
  10^{-8}$~\moyr.
  At a time $t\simeq$ 0.6 yr, the first working surfaces in both cones are
  formed.
  Initially, the flux densities increase, reaching maximum values of
  $S_\mathrm{1.3~cm}\simeq 0.078$ mJy and $S_\mathrm{3.6~cm}\simeq 0.085$ mJy
  at a time $t \simeq$ 0.85 yr.
  At this time, the model predicts a spectral index $\alpha_{1.3-3.6} \simeq$
  $-0.084$, as expected from optically thin emission.
  Afterward, the flux at both frequencies decreases until new working
  surfaces are formed.
  The predicted fluxes show a periodic behavior with the same oscillation
  period ($P$ = 1 yr) of the injected velocity.

 \begin{table}
   \caption{Predicted radio-continuum emission from proto-\ac{BD}
       candidates}

   \centering
   \addtolength{\tabcolsep}{1.85pt}

   \begin{tabular}{@{}lccccc@{}}
     \linss
     Model  &
     Geometry  &
     $V$\tnmk{a}  &
     $\dot m$\tnmk{b}  &
     $\nu$  &
     S$_{\nu}$\tnmk{c}  \\
     &
     &  
     (km s$^{-1}$)  &
     (M$_{\odot}$ yr$^{-1}$)  &
     (GHz)  &
     (mJy)  \\ 
     \linss
     C87\tnmk{d}  &  plane-parallel  &  100\tnmk{e}  &  2.0 $\times$
       10$^{-9}$  &  8.33  &  0.1 \\ 
     GC02\tnmk{f}  &  spherical  &  50--100\tnmk{g}  &  1.0 $\times$
       10$^{-8}$  &  8.33  &  0.083 \\ 
     R12\tnmk{h} & conical & 125\tnmk{i} & 5.0 $\times$ 10$^{-8}$  &  8.33  &
       0.085 \\ 
     \linss
   \end{tabular}

   \begin{minipage}{\columnwidth}
     \tntxt{a}{Shock velocities.}

     \tntxt{b}{Mass wind/outflow rate.}

     \tntxt{c}{Optically thin regime.}

     \tntxt{d}{\protect{\citet{Curiel87}} } 

     \tntxt{e}{Shock velocity (see \protect{\autoref{sec:model0}}).}

     \tntxt{f}{\protect{\citet{GonzalezCanto02}} }

     \tntxt{g}{\hang External and internal shock velocities, respectively (see
       \protect{\autoref{sec:model1}}).}

     \tntxt{h}{\protect{\citet{Rodriguez12}} }

     \tntxt{i}{\hang Mean velocity (see \protect{\autoref{sec:model2}}).}
   \end{minipage}

 \label{tab:continuum}
 \end{table}

\subsection{Proto-BDs as a scaled-down version of low-mass stars}
 \label{sec:proto_summary}

 \begin{figure}
   \centering
   \includegraphics[width=\columnwidth]{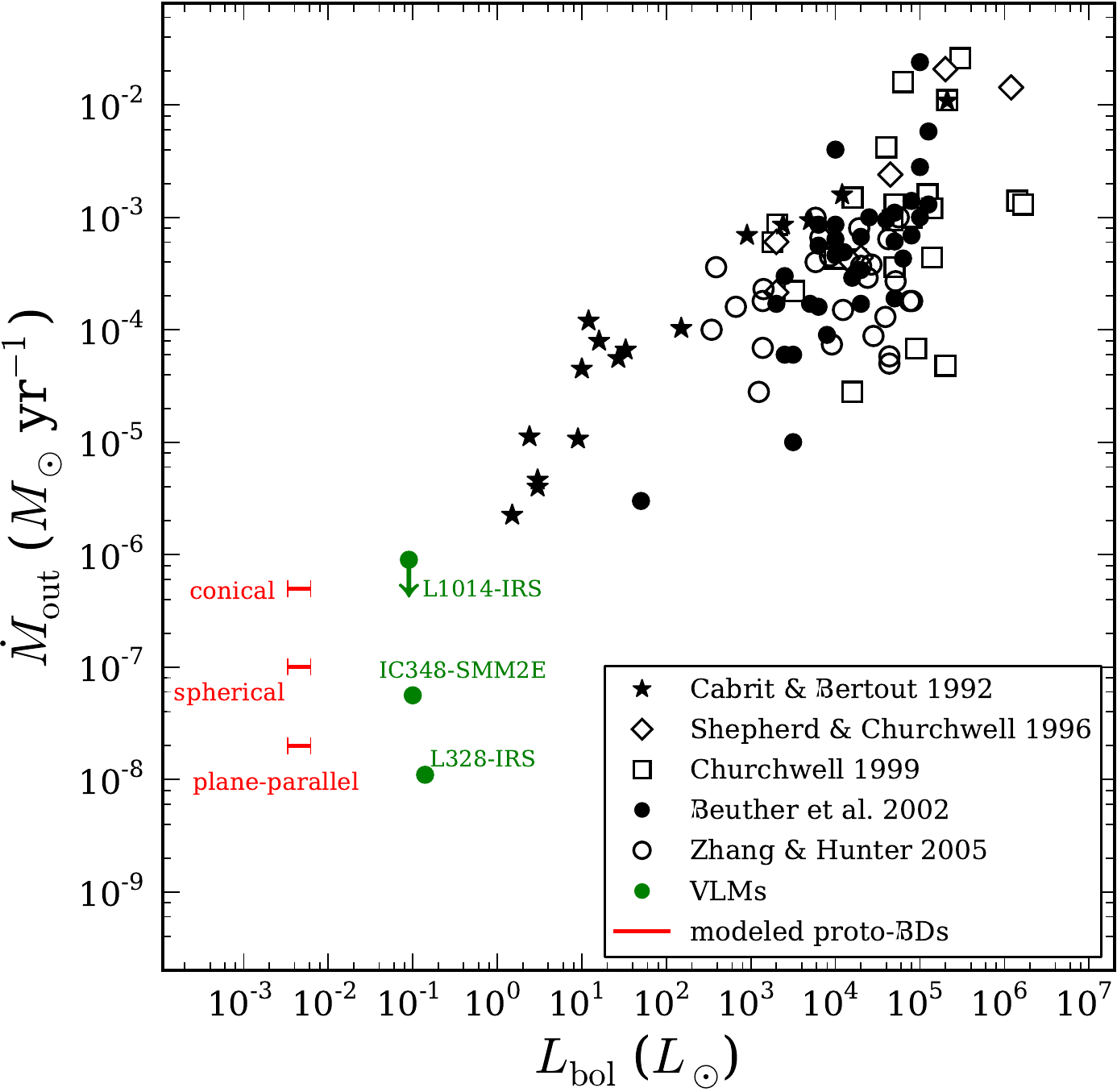}

   \caption{
     Mass outflow rate, $\mdotout$, vs.\ bolometric luminosity, $\Lbol$, for a
     sample of \acp{YSO} taken from: \protect{\citet{CabritBertout92}} (filled
     stars), \protect{\citet{Shepherd1996}} (open diamonds),
     \protect{\citet{Churchwell1999}} (open squares),
     \protect{\citet{Beuther2002}} (filled circles), and
     \protect{\citet{Zhang2005}} (open circles).
     The green dots indicate the values for three very low mass (VLMs)
     objects: a \ac{VeLLO}, L1014-IRS \citet{Shirley07}, and two very young
     proto-\acp{BD}, IC348-SMM2E and L328-IRS \citep{Lee13,Palau2014}.
     The horizontal red bars indicate the values for the $\Lbol$ of the
     proto-\ac{BD} candidates powering the detected thermal radio jets and the
     three different modeled values of $\mdotout$ (see \autoref{sec:models}).
   }
 \label{fig:OutvsBol}
 \end{figure}

  The models discussed in \autoref{sec:models} allowed us to explore different
  possible geometries for the thermal emission detected in our sample of
  proto-\ac{BD} candidates.
  In \autoref{tab:continuum}, we summarize the parameters used in the models
  to reproduce the observed centimeter emission, which range, for the mass
  loss rate, $\dot m$, from $2\times10^{-9}$ to
  $5\times10^{-8}$~\mo~yr$^{-1}$.
  Furthermore, we estimate a mass outflow rate, $\mdotout$, of the outflow
  that would be powered by this wind/conical outflow of about an order of
  magnitude higher, following \eg\ estimates from \citet{Beuther2002} and our
  own estimates from the outflow parameters obtained for the deeply embedded
  Class~0/I proto-\ac{BD} candidates IC348-SMM2E and L328-IRS
  \citep{Lee13,Palau2014}.
  We show in \autoref{fig:OutvsBol} how the mass outflow rates derived from
  the models lie in a typical relation found for young stellar objects for the
  mass outflow rate vs.\ the bolometric luminosity.
  We selected from the literature values corresponding to \acp{YSO} for a
  large range of $\Lbol$ and mass outflow rates.
  We used the values originally published by \citet{CabritBertout92},
  \citet{Shepherd1996}, \citet{Churchwell1999}, \citet{Beuther2002}, and
  \citet{Zhang2005}, and added the values of three very low mass objects (in
  green): the \ac{VeLLO} L1014-IRS, and the proto-\acp{BD} IC348-SMM2E and
  L328-IRS.
  \autoref{fig:OutvsBol} shows that the values inferred from our models are in
  agreement with the values expected if the trends found for young stellar
  objects are extrapolated to the luminosities of our proto-\ac{BD}
  candidates.
  In fact, the conical model seems to achieve the best agreement with the
  extrapolation, and reproduces the observed radio continuum fluxes only
  needing a relatively large mass outflow rate compared to the ones calculated
  for the two proto-BD candidates IC348-SMM2E and L328-IRS.
  Nonetheless, this difference in mass outflow rate is of about the same order
  of magnitude as the typical variation found for $\mdotout$ along the whole
  range of $\Lbol$, which can easily be of 1--2 orders of magnitude.
  However, since a spherically symmetric wind can also explain the radio
  observations with lower mass outflow rates, we believe that further research
  is needed to establish which mechanism is operating or even if a combination
  of mechanisms is needed.

  Thus, the models we used to explain the centimeter ``excess emission'' found
  in the $\Lcm$ vs.\ $\Lbol$ plot (\autoref{fig:LcmLbol}) require a range of
  mass outflow rates consistent with the values we would expect to find at
  very low bolometric luminosities if we would follow the trends found for
  \acp{YSO}.
  The models used in \autoref{sec:models} are the same ones that are usually
  applied to low-mass \acp{YSO}, but with physical parameters adapted to the
  \ac{BD} regime: smaller stellar wind velocities and smaller mass-loss rates
  than in low-mass \acp{YSO}.
  This supports the idea that the properties of proto-\acp{BD} are a
  continuation of the intrinsic properties of low-mass \acp{YSO} into the
  \ac{BD} regime.
  There is also some evidence, although still confined to a few star forming
  clouds and relatively small number of objects, that the presence of thermal
  radio jets may be more common in the Class~0 phase of star formation,
  compared to Class~I \citep[see][]{AMI2011a,AMI2011b,AMI2012}.
  As we mentioned in \autoref{sec:sed}, three of the four objects where we
  detect thermal radio jets are probably Class~0/I or young Class~I sources,
  given their relatively low $\Tbol$ values.
  These three sources are also among the four sources with the lower $\Tbol$
  in the sample, which is what we would expect to find if the detection ratio
  in Class~0 and Class~I \acp{YSO} is also extrapolated to the \ac{BD} regime.

  In order to explain the excess of centimeter emission with respect to the
  emission from \acp{YSO}, we propose that the initial pre-shock density could
  be be higher than what we assumed in our models, because proto-\acp{BD}
  could be less efficient in creating outflow cavities as compared to
  \acp{YSO}.
  This would also increase the efficiency of the shock and yield larger radio
  continuum fluxes.
  Additionally, the predictions of the models about the spectral indices of the
  centimeter emission are very consistent with our measurements, which are
  basically flat, indicating optically thin thermal emission.  
  All these results, plus the first detection of thermal radio jets in
  proto-\ac{BD} candidates, supports the idea that the formation mechanism of
  \acp{BD} is a scaled-down version of that of low-mass stars
  \citep{Chabrier2014}.

\section{Conclusions}
 \label{sec:conclusions}  

  We observed with the \ac{JVLA} at 3.6 and 1.3~cm a sample of 11
  proto-\ac{BD} candidates in Taurus, selected from \emph{Spitzer} data, that
  were associated with extended emission in 250~$\mu$m maps of
  \emph{Herschel}.
  We detected emission over 3$\sigma$ at both 1.3 and 3.6~cm in 5 sources
  (J041757, J041847, J041913, J041938, and J042123), only at 1.3~cm in
  J041836, and only at 3.6~cm in J041726 and J041740.
    Two of the sources, J041913 and J042123, are unresolved and clearly detected
  at both 3.6 and 1.3~cm, with intensity peaks between 0.3 and 0.9 m\jpb, and
  show negative spectral indices, $\sim-0.6$, tracing non-thermal synchrotron
  emission.
  We found using 21-cm continuum archival data that J042123 is a radiogalaxy
  and discarded it from further consideration.
  We also constructed the \acp{SED} for the remaining ten objects of our sample
  using data from near-IR to cm wavelengths and calculated their bolometric
  luminosities, $\Lbol$, and temperatures, $\Tbol$.
  We find that the bolometric temperatures of all the sources fall in the
  typical range of Class I protostellar objects.
  Globally, we detect a large fraction, 70\%, of our proto-\ac{BD} candidates
  at either 1.3 or 3.6 cm, which showcases the high sensitivity of the
  \ac{JVLA} for this kind of studies.

  Four of the 1.3-cm detected sources (J041757, J041836, J041847, and J041938)
  show partially extended, typically $\sim3''\times2''$, and faint
  emission $\sim0.1$~m\jpb, and their positions agree very well with the
  positions of the \emph{Spitzer} sources.
  The emission of these four sources shows flat or slightly positive spectral
  indices, consistent with optically thin or partially thick thermal free-free
  emission associated with a thermal radio jet emanating from the protostar, 
  which makes them excellent candidates to Class I proto-\acp{BD}.
  This would be the first detection of thermal radio jets in a sample of
  proto-\ac{BD} candidates.

  We also find that the four thermal radio jets show a centimeter emission
  excess when compared to the 3.6-cm vs.\ bolometric luminosity relationship
  found in low-mass \acp{YSO}.
  This excess is significant because $\Lbol$ is an upper limit to the true
  luminosity of the object and it is also difficult to explain invoking
  variable non-thermal emission or supersonic accretion onto a protostellar
  disk.
  We used models of the centimeter radio emission typically used for low-mass
  \acp{YSO} to explain this excess emission, assuming different
  geometries:
  a plane-parallel shock wave, a constant spherical stellar wind and a
  variable conical outflow; and adapting values for the mass loss rate and
  wind velocity to \acp{BD},
  $\dot{m}\sim2\times10^{-9}$--$5\times10^{-8}$\moyr and
  $v\sim50$--$100$~\kms.
  The models are able to reproduce approximately the measured fluxes at
  3.6~cm.
  We also find that the modeled mass outflow rates for the bolometric
  luminosities of our objects agree reasonably well with the trends found
  between $\mdotout$ and $\Lbol$ for \acp{YSO}.
  This indicates that the same mechanisms are at work for \acp{YSO} and
  proto-\acp{BD} and supports the idea that the intrinsic properties of
  proto-\acp{BD} are a continuation to smaller masses of the properties of
  low-mass \acp{YSO}.
  We propose that if \acp{BD} are less efficient in creating outflow cavities
  than \acp{YSO}, the initial pre-shock density could be higher and there
  would be an increase of the radio continuum flux, which would more easily
  explain the ``excess'' centimeter luminosities.

  These results, in addition to the detection of thermal radio jets and the
  \acp{SED} of our candidates consistent with Class I sources, suggests that
  the formation mechanism of our proto-\ac{BD} candidates is a scaled-down
  version of that of low-mass stars.
  Future observations will resolve the morphology of the radio jets associated
  with \acp{BD} and will test if they show less collimation than their higher
  mass counterparts.

\acknowledgements

  This work is based in part on data obtained as part of the UKIRT Infrared
  Deep Sky Survey.
  This publication makes use of data products from the Two Micron All Sky
  Survey, which is a joint project of the University of Massachusetts and the
  Infrared Processing and Analysis Center/California Institute of Technology,
  funded by the National Aeronautics and Space Administration and the National
  Science Foundation. 
  O.~Morata is supported by the MOST (Ministry of Science and Technology,
  Taiwan) ALMA-T grant MOST 103-2119-M-001-010-MY to the Institute of
  Astronomy \& Astrophysics, Academia Sinica.
  A.Palau acknowledges financial support from UNAM-DGAPA-PAPIIT IA102815
  grant, M\'exico. 
  I.~de Gregorio is supported by MICINN (Spain) grant AYA2011-30228-C03-01.
  R.~F.~Gonz\'alez is funded by UNAM/PAPIIT grant IN112014.
  H.~Bouy is funded by the the Ram\'on y Cajal fellowship program number
  RYC-2009-04497.
  H.~Bouy, D.~Barrado, N.~Hu\'elamo, and M.~Morales Calder\'on are supported
  by MICINN (Spain) grant AYA2012-38897-C02-01.
  A.~Bayo acknowledges financial support from the Proyecto Fondecyt de
  Iniciaci\'on 11140572.
  L.~F. Rodr\'{i}guez acknowledges the financial support from
  UNAM, and CONACyT, M\'exico.

\bibliographystyle{apj}
\bibliography{bds_evla.bbl}

\begin{thebibliography}{77}
\expandafter\ifx\csname natexlab\endcsname\relax\def\natexlab#1{#1}\fi

\bibitem[{{Alves de Oliveira} {et~al.}(2013){Alves de Oliveira}, {Moraux},
  {Bouvier}, {Duch{\^e}ne}, {Bouy}, {Maschberger}, \&
  {Hudelot}}]{AlvesOliveira13}
{Alves de Oliveira}, C., {Moraux}, E., {Bouvier}, J., {Duch{\^e}ne}, G.,
  {Bouy}, H., {Maschberger}, T., \& {Hudelot}, P. 2013, \aap, 549, A123

\bibitem[{{AMI Consortium} {et~al.}(2011{\natexlab{a}}){AMI Consortium},
  {Scaife}, {Curtis}, {Davies}, {Franzen}, {Grainge}, {Hobson},
  {Hurley-Walker}, {Lasenby}, {Olamaie}, {Pooley},
  {Rodr{\'{\i}}guez-Gonz{\'a}lvez}, {Saunders}, {Schammel}, {Scott},
  {Shimwell}, {Titterington}, {Waldram}, \& {Zwart}}]{AMI2011a}
{AMI Consortium} {et~al.} 2011{\natexlab{a}}, \mnras, 410, 2662

\bibitem[{{AMI Consortium} {et~al.}(2011{\natexlab{b}}){AMI Consortium},
  {Scaife}, {Hatchell}, {Davies}, {Franzen}, {Grainge}, {Hobson},
  {Hurley-Walker}, {Lasenby}, {Olamaie}, {Perrott}, {Pooley},
  {Rodr{\'{\i}}guez-Gonz{\'a}lvez}, {Saunders}, {Schammel}, {Scott},
  {Shimwell}, {Titterington}, \& {Waldram}}]{AMI2011b}
---. 2011{\natexlab{b}}, \mnras, 415, 893

\bibitem[{{AMI Consortium} {et~al.}(2012){AMI Consortium}, {Scaife},
  {Hatchell}, {Davies}, {Franzen}, {Grainge}, {Hobson}, {Hurley-Walker},
  {Lasenby}, {Olamaie}, {Perrott}, {Pooley}, {Rodr{\'{\i}}guez-Gonz{\'a}lvez},
  {Saunders}, {Schammel}, {Scott}, {Shimwell}, {Titterington}, \&
  {Waldram}}]{AMI2012}
---. 2012, \mnras, 420, 1019

\bibitem[{{Andr{\'e}} {et~al.}(2010){Andr{\'e}}, {Men'shchikov}, {Bontemps},
  {K{\"o}nyves}, {Motte}, {Schneider}, {Didelon}, {Minier}, {Saraceno},
  {Ward-Thompson}, {di Francesco}, {White}, {Molinari}, {Testi}, {Abergel},
  {Griffin}, {Henning}, {Royer}, {Mer{\'{\i}}n}, {Vavrek}, {Attard},
  {Arzoumanian}, {Wilson}, {Ade}, {Aussel}, {Baluteau}, {Benedettini},
  {Bernard}, {Blommaert}, {Cambr{\'e}sy}, {Cox}, {di Giorgio}, {Hargrave},
  {Hennemann}, {Huang}, {Kirk}, {Krause}, {Launhardt}, {Leeks}, {Le Pennec},
  {Li}, {Martin}, {Maury}, {Olofsson}, {Omont}, {Peretto}, {Pezzuto}, {Prusti},
  {Roussel}, {Russeil}, {Sauvage}, {Sibthorpe}, {Sicilia-Aguilar}, {Spinoglio},
  {Waelkens}, {Woodcraft}, \& {Zavagno}}]{Andre10}
{Andr{\'e}}, P., {et~al.} 2010, \aap, 518, L102

\bibitem[{{Andr{\'e}} {et~al.}(1999){Andr{\'e}}, {Motte}, \&
  {Bacmann}}]{Andre99}
{Andr{\'e}}, P., {Motte}, F., \& {Bacmann}, A. 1999, \apjl, 513, L57

\bibitem[{{Andr{\'e}} {et~al.}(2012){Andr{\'e}}, {Ward-Thompson}, \&
  {Greaves}}]{Andre2012}
{Andr{\'e}}, P., {Ward-Thompson}, D., \& {Greaves}, J. 2012, Science, 337, 69

\bibitem[{Andr\'e {et~al.}(1993)Andr\'e, Ward-Thompson, \& M.}]{Andre93}
Andr\'e, P., Ward-Thompson, D., \& M., B. 1993, ApJ, 406, 122

\bibitem[{{Anglada}(1995)}]{Anglada95}
{Anglada}, G. 1995, in Revista Mexicana de Astronomia y Astrofisica Conference
  Series, Vol.~1, Revista Mexicana de Astronomia y Astrofisica Conference
  Series, ed. S.~{Lizano} \& J.~M. {Torrelles}, 67

\bibitem[{{Bally} {et~al.}(2007){Bally}, {Reipurth}, \& {Davis}}]{Bally2007}
{Bally}, J., {Reipurth}, B., \& {Davis}, C.~J. 2007, Protostars and Planets V,
  215

\bibitem[{{Baraffe} {et~al.}(2002){Baraffe}, {Chabrier}, {Allard}, \&
  {Hauschildt}}]{Baraffe02}
{Baraffe}, I., {Chabrier}, G., {Allard}, F., \& {Hauschildt}, P.~H. 2002, \aap,
  382, 563

\bibitem[{{Barrado} {et~al.}(2009){Barrado}, {Morales-Calder{\'o}n}, {Palau},
  {Bayo}, {de Gregorio-Monsalvo}, {Eiroa}, {Hu{\'e}lamo}, {Bouy}, {Morata}, \&
  {Schmidtobreick}}]{Barrado09}
{Barrado}, D., {et~al.} 2009, \aap, 508, 859

\bibitem[{{Bate}(2012)}]{Bate12}
{Bate}, M.~R. 2012, \mnras, 419, 3115

\bibitem[{{Bayo} {et~al.}(2012){Bayo}, {Barrado}, {Hu{\'e}lamo},
  {Morales-Calder{\'o}n}, {Melo}, {Stauffer}, \& {Stelzer}}]{Bayo12}
{Bayo}, A., {Barrado}, D., {Hu{\'e}lamo}, N., {Morales-Calder{\'o}n}, M.,
  {Melo}, C., {Stauffer}, J., \& {Stelzer}, B. 2012, \aap, 547, A80

\bibitem[{{Bayo} {et~al.}(2011){Bayo}, {Barrado}, {Stauffer},
  {Morales-Calder{\'o}n}, {Melo}, {Hu{\'e}lamo}, {Bouy}, {Stelzer}, {Tamura},
  \& {Jayawardhana}}]{Bayo11}
{Bayo}, A., {et~al.} 2011, \aap, 536, A63

\bibitem[{{Beltr{\'a}n} {et~al.}(2001){Beltr{\'a}n}, {Estalella}, {Anglada},
  {Rodr{\'{\i}}guez}, \& {Torrelles}}]{Beltran01}
{Beltr{\'a}n}, M.~T., {Estalella}, R., {Anglada}, G., {Rodr{\'{\i}}guez},
  L.~F., \& {Torrelles}, J.~M. 2001, \aj, 121, 1556

\bibitem[{{Beuther} {et~al.}(2002){Beuther}, {Schilke}, {Sridharan}, {Menten},
  {Walmsley}, \& {Wyrowski}}]{Beuther2002}
{Beuther}, H., {Schilke}, P., {Sridharan}, T.~K., {Menten}, K.~M., {Walmsley},
  C.~M., \& {Wyrowski}, F. 2002, \aap, 383, 892

\bibitem[{{Bieging} \& {Cohen}(1989)}]{BiegingCohen89}
{Bieging}, J.~H., \& {Cohen}, M. 1989, \aj, 98, 1686

\bibitem[{{Bontemps} {et~al.}(1996){Bontemps}, {Andre}, {Terebey}, \&
  {Cabrit}}]{Bontemps96}
{Bontemps}, S., {Andre}, P., {Terebey}, S., \& {Cabrit}, S. 1996, \aap, 311,
  858

\bibitem[{{Cabrit} \& {Bertout}(1992)}]{CabritBertout92}
{Cabrit}, S., \& {Bertout}, C. 1992, \aap, 261, 274

\bibitem[{{Chabrier} {et~al.}(2014){Chabrier}, {Johansen}, {Janson}, \&
  {Rafikov}}]{Chabrier2014}
{Chabrier}, G., {Johansen}, A., {Janson}, M., \& {Rafikov}, R. 2014, Protostars
  and Planets VI, 619

\bibitem[{{Chen} {et~al.}(1995){Chen}, {Myers}, {Ladd}, \& {Wood}}]{Chen95}
{Chen}, H., {Myers}, P.~C., {Ladd}, E.~F., \& {Wood}, D.~O.~S. 1995, \apj, 445,
  377

\bibitem[{{Choi} {et~al.}(2014){Choi}, {Lee}, \& {Kang}}]{Choi14}
{Choi}, M., {Lee}, J.-E., \& {Kang}, M. 2014, \apj, 789, 9

\bibitem[{{Churchwell}(1999)}]{Churchwell1999}
{Churchwell}, E. 1999, in NATO Advanced Science Institutes (ASI) Series C, Vol.
  540, NATO Advanced Science Institutes (ASI) Series C, ed. C.~J. {Lada} \&
  N.~D. {Kylafis}, 515

\bibitem[{{Condon} {et~al.}(1998){Condon}, {Cotton}, {Greisen}, {Yin},
  {Perley}, {Taylor}, \& {Broderick}}]{NVSS98}
{Condon}, J.~J., {Cotton}, W.~D., {Greisen}, E.~W., {Yin}, Q.~F., {Perley},
  R.~A., {Taylor}, G.~B., \& {Broderick}, J.~J. 1998, \aj, 115, 1693

\bibitem[{{Curiel} {et~al.}(1987){Curiel}, {Canto}, \& {Rodriguez}}]{Curiel87}
{Curiel}, S., {Canto}, J., \& {Rodriguez}, L.~F. 1987, RMxAA, 14, 595

\bibitem[{{Curiel} {et~al.}(1989){Curiel}, {Rodriguez}, {Bohigas}, {Roth},
  {Canto}, \& {Torrelles}}]{Curiel89}
{Curiel}, S., {Rodriguez}, L.~F., {Bohigas}, J., {Roth}, M., {Canto}, J., \&
  {Torrelles}, J.~M. 1989, Astrophysical Letters and Communications, 27, 299

\bibitem[{{Di Francesco} {et~al.}(2007){Di Francesco}, {Evans}, {Caselli},
  {Myers}, {Shirley}, {Aikawa}, \& {Tafalla}}]{DiFrancesco07}
{Di Francesco}, J., {Evans}, II, N.~J., {Caselli}, P., {Myers}, P.~C.,
  {Shirley}, Y., {Aikawa}, Y., \& {Tafalla}, M. 2007, Protostars and Planets V,
  17

\bibitem[{{Dzib} {et~al.}(2013){Dzib}, {Loinard}, {Mioduszewski},
  {Rodr{\'{\i}}guez}, {Ortiz-Le{\'o}n}, {Pech}, {Rivera}, {Torres}, {Boden},
  {Hartmann}, {Evans}, {Brice{\~n}o}, \& {Tobin}}]{Dzib13}
{Dzib}, S.~A., {et~al.} 2013, \apj, 775, 63

\bibitem[{{Frank} {et~al.}(2014){Frank}, {Ray}, {Cabrit}, {Hartigan}, {Arce},
  {Bacciotti}, {Bally}, {Benisty}, {Eisl{\"o}ffel}, {G{\"u}del}, {Lebedev},
  {Nisini}, \& {Raga}}]{Frank2014}
{Frank}, A., {et~al.} 2014, Protostars and Planets VI, 451

\bibitem[{{Furuya} {et~al.}(2003){Furuya}, {Kitamura}, {Wootten}, {Claussen},
  \& {Kawabe}}]{Furuya03}
{Furuya}, R.~S., {Kitamura}, Y., {Wootten}, A., {Claussen}, M.~J., \& {Kawabe},
  R. 2003, \apjs, 144, 71

\bibitem[{{Girart} {et~al.}(2002){Girart}, {Curiel}, {Rodr{\'{\i}}guez}, \&
  {Cant{\'o}}}]{Girart02}
{Girart}, J.~M., {Curiel}, S., {Rodr{\'{\i}}guez}, L.~F., \& {Cant{\'o}}, J.
  2002, RMxAA, 38, 169

\bibitem[{{Gonz{\'a}lez}(2002)}]{Gonzalez02}
{Gonz{\'a}lez}, R.~F. 2002, PhD thesis, Univ.\ Nacional Aut{\'o}noma de
  M{\'e}xico

\bibitem[{{Gonz{\'a}lez} \& {Cant{\'o}}(2002)}]{GonzalezCanto02}
{Gonz{\'a}lez}, R.~F., \& {Cant{\'o}}, J. 2002, \apj, 580, 459

\bibitem[{{Gonz{\'a}lez} \& {Cant{\'o}}(2008)}]{GonzalezCanto08}
---. 2008, \aap, 477, 373

\bibitem[{{Joergens} {et~al.}(2013){Joergens}, {Bonnefoy}, {Liu}, {Bayo},
  {Wolf}, {Chauvin}, \& {Rojo}}]{Joergens13}
{Joergens}, V., {Bonnefoy}, M., {Liu}, Y., {Bayo}, A., {Wolf}, S., {Chauvin},
  G., \& {Rojo}, P. 2013, \aap, 558, L7

\bibitem[{{Joergens} {et~al.}(2012){Joergens}, {Pohl}, {Sicilia-Aguilar}, \&
  {Henning}}]{Joergens12}
{Joergens}, V., {Pohl}, A., {Sicilia-Aguilar}, A., \& {Henning}, T. 2012, \aap,
  543, A151

\bibitem[{{Kang} \& {Shapiro}(1992)}]{Kang1992}
{Kang}, H., \& {Shapiro}, P.~R. 1992, \apj, 386, 432

\bibitem[{{Kraus}(1986)}]{Kraus1986}
{Kraus}, J.~D. 1986, {Radio Astronomy}, 2nd edn. (Cygnus-Quasar Books)

\bibitem[{{Kroupa} \& {Bouvier}(2003)}]{Kroupa2003}
{Kroupa}, P., \& {Bouvier}, J. 2003, \mnras, 346, 369

\bibitem[{Lada(1985)}]{Lada85}
Lada, C.~J. 1985, ARA\&A, 23, 267

\bibitem[{{Lee} {et~al.}(2013){Lee}, {Kim}, {Kim}, {Saito}, {Myers}, \&
  {Kurono}}]{Lee13}
{Lee}, C.~W., {Kim}, M.-R., {Kim}, G., {Saito}, M., {Myers}, P.~C., \&
  {Kurono}, Y. 2013, \apj, 777, 50

\bibitem[{{Li} {et~al.}(2014){Li}, {Banerjee}, {Pudritz}, {J{\o}rgensen},
  {Shang}, {Krasnopolsky}, \& {Maury}}]{Li2014}
{Li}, Z.-Y., {Banerjee}, R., {Pudritz}, R.~E., {J{\o}rgensen}, J.~K., {Shang},
  H., {Krasnopolsky}, R., \& {Maury}, A. 2014, Protostars and Planets VI, 173

\bibitem[{{Loinard} {et~al.}(2005){Loinard}, {Mioduszewski},
  {Rodr{\'{\i}}guez}, {Gonz{\'a}lez}, {Rodr{\'{\i}}guez}, \&
  {Torres}}]{Loinard05}
{Loinard}, L., {Mioduszewski}, A.~J., {Rodr{\'{\i}}guez}, L.~F.,
  {Gonz{\'a}lez}, R.~A., {Rodr{\'{\i}}guez}, M.~I., \& {Torres}, R.~M. 2005,
  \apjl, 619, L179

\bibitem[{{Luhman}(2012)}]{Luhman2012}
{Luhman}, K.~L. 2012, \araa, 50, 65

\bibitem[{{Lynch} {et~al.}(2013){Lynch}, {Mutel}, {G{\"u}del}, {Ray},
  {Skinner}, {Schneider}, \& {Gayley}}]{Lynch13}
{Lynch}, C., {Mutel}, R.~L., {G{\"u}del}, M., {Ray}, T., {Skinner}, S.~L.,
  {Schneider}, P.~C., \& {Gayley}, K.~G. 2013, \apj, 766, 53

\bibitem[{{McMullin} {et~al.}(2007){McMullin}, {Waters}, {Schiebel}, {Young},
  \& {Golap}}]{casaref}
{McMullin}, J.~P., {Waters}, B., {Schiebel}, D., {Young}, W., \& {Golap}, K.
  2007, in Astronomical Society of the Pacific Conference Series, Vol. 376,
  Astronomical Data Analysis Software and Systems XVI, ed. R.~A. {Shaw},
  F.~{Hill}, \& D.~J. {Bell}, 127

\bibitem[{{Monin} {et~al.}(2013){Monin}, {Whelan}, {Lefloch}, {Dougados}, \&
  {Alves de Oliveira}}]{Monin2013}
{Monin}, J.-L., {Whelan}, E.~T., {Lefloch}, B., {Dougados}, C., \& {Alves de
  Oliveira}, C. 2013, \aap, 551, L1

\bibitem[{{Mu{\v z}i{\'c}} {et~al.}(2014){Mu{\v z}i{\'c}}, {Scholz}, {Geers},
  {Jayawardhana}, \& {L{\'o}pez Mart{\'{\i}}}}]{Muzic14}
{Mu{\v z}i{\'c}}, K., {Scholz}, A., {Geers}, V.~C., {Jayawardhana}, R., \&
  {L{\'o}pez Mart{\'{\i}}}, B. 2014, \apj, 785, 159

\bibitem[{Neufeld \& Hollenbach(1996)}]{NeufeldHollenbach96}
Neufeld, D.~A., \& Hollenbach, D.~J. 1996, \apjl, 471, L45

\bibitem[{{Ott}(2010)}]{Ott10}
{Ott}, S. 2010, in Astronomical Society of the Pacific Conference Series, Vol.
  434, Astronomical Data Analysis Software and Systems XIX, ed. Y.~{Mizumoto},
  K.-I. {Morita}, \& M.~{Ohishi}, 139

\bibitem[{{Palau} {et~al.}(2012){Palau}, {de Gregorio-Monsalvo}, {Morata},
  {Stamatellos}, {Hu{\'e}lamo}, {Eiroa}, {Bayo}, {Morales-Calder{\'o}n},
  {Bouy}, {Ribas}, {Asmus}, \& {Barrado}}]{Palau12}
{Palau}, A., {et~al.} 2012, \mnras, 424, 2778

\bibitem[{{Palau} {et~al.}(2014){Palau}, {Zapata}, {Rodr{\'{\i}}guez}, {Bouy},
  {Barrado}, {Morales-Calder{\'o}n}, {Myers}, {Chapman}, {Ju{\'a}rez}, \&
  {Li}}]{Palau2014}
---. 2014, \mnras, 444, 833

\bibitem[{{Phan-Bao} {et~al.}(2014{\natexlab{a}}){Phan-Bao}, {Lee}, {Ho},
  {Dang-Duc}, \& {Li}}]{PhanBao2014a}
{Phan-Bao}, N., {Lee}, C.-F., {Ho}, P.~T.~P., {Dang-Duc}, C., \& {Li}, D.
  2014{\natexlab{a}}, \apj, 795, 70

\bibitem[{{Phan-Bao} {et~al.}(2014{\natexlab{b}}){Phan-Bao}, {Lee}, {Ho}, \&
  {Mart{\'{\i}}n}}]{PhanBao14}
{Phan-Bao}, N., {Lee}, C.-F., {Ho}, P.~T.~P., \& {Mart{\'{\i}}n}, E.~L.
  2014{\natexlab{b}}, \aap, 564, A32

\bibitem[{{Phan-Bao} {et~al.}(2011){Phan-Bao}, {Lee}, {Ho}, \&
  {Tang}}]{PhanBao2011}
{Phan-Bao}, N., {Lee}, C.-F., {Ho}, P.~T.~P., \& {Tang}, Y.-W. 2011, \apj, 735,
  14

\bibitem[{{Phan-Bao} {et~al.}(2008){Phan-Bao}, {Riaz}, {Lee}, {Tang}, {Ho},
  {Mart{\'{\i}}n}, {Lim}, {Ohashi}, \& {Shang}}]{PhanBao2008}
{Phan-Bao}, N., {et~al.} 2008, \apjl, 689, L141

\bibitem[{{Ray} {et~al.}(2007){Ray}, {Dougados}, {Bacciotti}, {Eisl{\"o}ffel},
  \& {Chrysostomou}}]{Ray2007}
{Ray}, T., {Dougados}, C., {Bacciotti}, F., {Eisl{\"o}ffel}, J., \&
  {Chrysostomou}, A. 2007, Protostars and Planets V, 231

\bibitem[{{Reipurth} \& {Clarke}(2001)}]{Reipurth2001}
{Reipurth}, B., \& {Clarke}, C. 2001, \aj, 122, 432

\bibitem[{{Reipurth} {et~al.}(2002){Reipurth}, {Rodr{\'{\i}}guez}, {Anglada},
  \& {Bally}}]{Reipurth02}
{Reipurth}, B., {Rodr{\'{\i}}guez}, L.~F., {Anglada}, G., \& {Bally}, J. 2002,
  \aj, 124, 1045

\bibitem[{{Riaz} {et~al.}(2015){Riaz}, {Thompson}, {Whelan}, \&
  {Lodieu}}]{Riaz2015}
{Riaz}, B., {Thompson}, M., {Whelan}, E.~T., \& {Lodieu}, N. 2015, \mnras, 446,
  2550

\bibitem[{{Rodr\'{i}guez}(1998)}]{Rodriguez98}
{Rodr\'{i}guez}, L.~F. 1998, in Revista Mexicana de Astronomia y Astrofisica
  Conference Series, Vol.~7, Revista Mexicana de Astronomia y Astrofisica
  Conference Series, ed. R.~J. {Dufour} \& S.~{Torres-Peimbert}, 14--20

\bibitem[{{Rodr{\'{\i}}guez} {et~al.}(2012){Rodr{\'{\i}}guez}, {Gonz{\'a}lez},
  {Raga}, {Cant{\'o}}, {Riera}, {Loinard}, {Dzib}, \& {Zapata}}]{Rodriguez12}
{Rodr{\'{\i}}guez}, L.~F., {Gonz{\'a}lez}, R.~F., {Raga}, A.~C., {Cant{\'o}},
  J., {Riera}, A., {Loinard}, L., {Dzib}, S.~A., \& {Zapata}, L.~A. 2012, \aap,
  537, A123

\bibitem[{{Roussel}(2013)}]{Roussel13}
{Roussel}, H. 2013, \pasp, 125, 1126

\bibitem[{{Scholz} {et~al.}(2012){Scholz}, {Jayawardhana}, {Muzic}, {Geers},
  {Tamura}, \& {Tanaka}}]{Scholz12}
{Scholz}, A., {Jayawardhana}, R., {Muzic}, K., {Geers}, V., {Tamura}, M., \&
  {Tanaka}, I. 2012, \apj, 756, 24

\bibitem[{{Shepherd} \& {Churchwell}(1996)}]{Shepherd1996}
{Shepherd}, D.~S., \& {Churchwell}, E. 1996, \apj, 472, 225

\bibitem[{{Shirley} {et~al.}(2007){Shirley}, {Claussen}, {Bourke}, {Young}, \&
  {Blake}}]{Shirley07}
{Shirley}, Y.~L., {Claussen}, M.~J., {Bourke}, T.~L., {Young}, C.~H., \&
  {Blake}, G.~A. 2007, \apj, 667, 329

\bibitem[{{Skrutskie} {et~al.}(2006){Skrutskie}, {Cutri}, {Stiening},
  {Weinberg}, {Schneider}, {Carpenter}, {Beichman}, {Capps}, {Chester},
  {Elias}, {Huchra}, {Liebert}, {Lonsdale}, {Monet}, {Price}, {Seitzer},
  {Jarrett}, {Kirkpatrick}, {Gizis}, {Howard}, {Evans}, {Fowler}, {Fullmer},
  {Hurt}, {Light}, {Kopan}, {Marsh}, {McCallon}, {Tam}, {Van Dyk}, \&
  {Wheelock}}]{Skrutskie06}
{Skrutskie}, M.~F., {et~al.} 2006, \aj, 131, 1163

\bibitem[{{Stamatellos} \& {Whitworth}(2009)}]{StamatellosWhitworth09}
{Stamatellos}, D., \& {Whitworth}, A.~P. 2009, \mnras, 392, 413

\bibitem[{{Umbreit} {et~al.}(2005){Umbreit}, {Burkert}, {Henning}, {Mikkola},
  \& {Spurzem}}]{Umbreit2005}
{Umbreit}, S., {Burkert}, A., {Henning}, T., {Mikkola}, S., \& {Spurzem}, R.
  2005, \apj, 623, 940

\bibitem[{{Whelan} {et~al.}(2014){Whelan}, {Alcal{\'a}}, {Bacciotti}, {Nisini},
  {Bonito}, {Antoniucci}, {Stelzer}, {Biazzo}, {D'Elia}, \&
  {Ray}}]{Whelan2014b}
{Whelan}, E.~T., {et~al.} 2014, \aap, 570, A59

\bibitem[{{Whelan} {et~al.}(2009{\natexlab{a}}){Whelan}, {Ray}, \&
  {Bacciotti}}]{Whelan2009a}
{Whelan}, E.~T., {Ray}, T.~P., \& {Bacciotti}, F. 2009{\natexlab{a}}, \apjl,
  691, L106

\bibitem[{{Whelan} {et~al.}(2005){Whelan}, {Ray}, {Bacciotti}, {Natta},
  {Testi}, \& {Randich}}]{Whelan05}
{Whelan}, E.~T., {Ray}, T.~P., {Bacciotti}, F., {Natta}, A., {Testi}, L., \&
  {Randich}, S. 2005, \nat, 435, 652

\bibitem[{{Whelan} {et~al.}(2012){Whelan}, {Ray}, {Comeron}, {Bacciotti}, \&
  {Kavanagh}}]{Whelan12}
{Whelan}, E.~T., {Ray}, T.~P., {Comeron}, F., {Bacciotti}, F., \& {Kavanagh},
  P.~J. 2012, \apj, 761, 120

\bibitem[{{Whelan} {et~al.}(2009{\natexlab{b}}){Whelan}, {Ray}, {Podio},
  {Bacciotti}, \& {Randich}}]{Whelan2009b}
{Whelan}, E.~T., {Ray}, T.~P., {Podio}, L., {Bacciotti}, F., \& {Randich}, S.
  2009{\natexlab{b}}, \apj, 706, 1054

\bibitem[{{Whitworth} {et~al.}(2007){Whitworth}, {Bate}, {Nordlund},
  {Reipurth}, \& {Zinnecker}}]{Whitworth2007}
{Whitworth}, A., {Bate}, M.~R., {Nordlund}, {\AA}., {Reipurth}, B., \&
  {Zinnecker}, H. 2007, Protostars and Planets V, 459

\bibitem[{{Zhang} {et~al.}(2005){Zhang}, {Hunter}, {Brand}, {Sridharan},
  {Cesaroni}, {Molinari}, {Wang}, \& {Kramer}}]{Zhang2005}
{Zhang}, Q., {Hunter}, T.~R., {Brand}, J., {Sridharan}, T.~K., {Cesaroni}, R.,
  {Molinari}, S., {Wang}, J., \& {Kramer}, M. 2005, \apj, 625, 864

\end{thebibliography}

\appendix

\section{Photometric data for Spectral Energy Distributions}
 \label{apx:sed_data}

 \begin{table}[ht]
   \caption{Photometry for J041726}
   \centering
     {\small
       \begin{tabular}{ccccl}
         \linss
         $\lambda$  &  $S_\nu$  &  $\sigma_\mathrm{abs}$\tnmk{a}  & Beam \\
         ($\mu$m)  &  (mJy)  &  (mJy)  &  (arcsec)  &  Instrument \\
         \linss
         3.4  &  0.477  &  0.018  &  2.3  &  WISE\\
         3.6  &  0.399  &  0.009  &  1.7  &  Spitzer/IRAC\\
         4.5  &  0.486  &  0.015  &  1.7  &  Spitzer/IRAC\\
         4.6  &  0.608  &  0.029  &  2.9  &  WISE\\
         5.8  &  0.521  &  0.039  &  1.9  &  Spitzer/IRAC\\
         8.0  &  0.944  &  0.054  &  2.0  &  Spitzer/IRAC\\
         12  &  1.60  &  0.17  &  7.6  &  WISE\\
         22  &  6.64  &  1.10  &  13.8  &  WISE\\
         70		&$<18$ 	 	&$-$  &  5.6  &  Herschel/PACS\\
         160		&$<140$  		&$-$  &  11  &  Herschel/PACS\\
         250		&$<576$  		&$-$  &  18  &  Herschel/SPIRE\\
         350		&$<415$  		&$-$  &  25  &  Herschel/SPIRE\\
         500		&$<480$  		&$-$  &  35  &  Herschel/SPIRE\\
         1200		&$<4.2$		&$-$  &  11  &  IRAM30m/MAMBO\\
         13000 &  $<0.036$  &  $-$  &  $1.5\times1.2$  &  JVLA\\
         36000 &  0.04  &  0.02  &  $2.5\times 1.8$  & JVLA\\
         \linss
       \end{tabular}

       \begin{minipage}{8cm}
         \tntxt{a}{Absolute flux uncertainty.
           We adopted an absolute flux uncertainty of 25\% for \emph{Herschel}
           measurements.
         }
       \end{minipage}
     }
 \label{tab:app_sed_J041726}
 \end{table}

 \begin{table}
   \caption{Photometry for J041740}
   \centering
     {\small
       \begin{tabular}{ccccl}
         \linss
         $\lambda$
         &$S_\nu$
         &$\sigma_\mathrm{abs}$\tnmk{a}
         &Beam
         \\
         ($\mu$m)
         &(mJy)
         &(mJy)
         &(arcsec)
         &Instrument
         \\
         \linss
         1.23  &  0.175    &  0.017  &  2.5	  &  2MASS\\
         1.66  &  0.423    &  0.042  &  2.5	  &  2MASS\\
         2.16  &  0.631    &  0.071  &  2.5	  &  2MASS\\
         3.4  &  0.533    &  0.016  &  2.3	  &  WISE\\
         3.6  &  0.615    &  0.014  &  1.7	  &  Spitzer/IRAC\\
         4.5  &  0.584    &  0.014  &  1.7	  &  Spitzer/IRAC\\
         4.6  &  0.565    &  0.022  &  2.9	  &  WISE\\
         5.8  &  0.347    &  0.045  &  1.9	  &  Spitzer/IRAC\\
         8.0  &  3.30    &  0.06  &  2.0	  &  Spitzer/IRAC\\
         12  &  3.27    &  0.24  &  7.6	  &  WISE\\
         22  &  3.65    &  1.10  &  13.8  &  WISE\\
         24  &  4.48    &  0.34  &  6.0	  &  Spitzer/MIPS\\
         70  &  18 	   &  4	  &  5.6	  &  Herschel/PACS\\
         160  &  104    &  26	  &  11	  &  Herschel/PACS\\
         250  &  62    &  16	  &  18	  &  Herschel/SPIRE\\
         350  &  14    &  4	  &  25	  &  Herschel/SPIRE\\
         500  &  $<245$    &  $-$	  &  35	  &  Herschel/SPIRE\\
         13000 &$<0.040$  &$-$   &  $1.5\times1.2$ &JVLA\\
         36000 &0.09   &0.03  &  $2.6\times 1.9$ &JVLA\\
         \linss
       \end{tabular}

       \begin{minipage}{7.5cm}
         \tntxt{a}{Absolute flux uncertainty.
           We adopted an absolute flux uncertainty of 25\% for \emph{Herschel}
           measurements.
         }
       \end{minipage}
     }
 \label{tab:app_sed_J041740}
 \end{table}

 \begin{table}
   \caption{Photometry for J041757 (component B of \citealt{Palau12}).}
   \centering
     {\small
       \begin{tabular}{ccccl}
         \linss
         $\lambda$
         &$S_\nu$
         &$\sigma_\mathrm{abs}$\tnmk{a}
         &Beam
         \\
         ($\mu$m)
         &(mJy)
         &(mJy)
         &(arcsec)
         &Instrument
         \\
         \linss
         0.75  &  0.0034  &  0.0001  &  1	  &  CFHT/MegaCam\\
         0.90  &  0.0077  &  0.0002  &  1	  &  CFHT/MegaCam\\
         1.03  &  $<0.130$  	&$-$	  &  0.6	  &  UKIDSS/WFCAM\\
         1.25  &  0.0360    &  0.0006  &  1	  &  CAHA/Omega2000\\
         1.65  &  0.0680    &  0.0012  &  1	  &  CAHA/Omega2000\\
         2.17  &  0.127    &  0.002  &  1	  &  CAHA/Omega2000\\
         3.4  &  0.435    &  0.017  &  2.3	  &  WISE\\
         3.6  &  0.295    &  0.008  &  1.7	  &  Spitzer/IRAC\\
         4.5  &  0.441    &  0.009  &  1.7	  &  Spitzer/IRAC\\
         4.6  &  0.725    &  0.028  &  2.9	  &  WISE\\
         5.8  &  0.491    &  0.010  &  1.9	  &  Spitzer/IRAC\\
         8.0  &  0.736    &  0.011  &  2.0	  &  Spitzer/IRAC\\
         12  &  1.79    &  0.15  &  7.6	  &  WISE\\
         22  &  6.5    &  1.0	  &  13.8  &  WISE\\
         24  &  7.5    &  1.0	  &  6.0	  &  Spitzer/MIPS\\
         70  &  $<23$ 	 	&$-$	  &  5.6	  &  Herschel/PACS\\
         160  &  57    &  14	  &  11	  &  Herschel/PACS\\
         250  &  86    &  21	  &  18	  &  Herschel/SPIRE\\
         350  &  $<169$    &  $-$	  &  25	  &  Herschel/SPIRE\\
         350  &  165  &  9	  &  10	  &  CSO/SHARC\\
         500  &  $<132$    &  $-$	  &  35	  &  Herschel/SPIRE\\
         870  &  $<80$  &  $-$	  &  18	  &  APEX/LABOCA\\
         1200  &  $<2.9$  &  $-$	  &  11	  &  IRAM30m/MAMBO\\
         13000 &0.13   &0.03  &  $2.0\times1.8$ &JVLA\\
         36000 &0.14   &0.04  &  $2.3\times 1.7$ &JVLA\\
         \linss
       \end{tabular}

       \begin{minipage}{8.25cm}
         \tntxt{a}{Absolute flux uncertainty.
           We adopted an absolute flux uncertainty of 25\% for \emph{Herschel}
           measurements.
         }
       \end{minipage}
     }

 \label{tab:app_sed_J041757}
 \end{table}

 \begin{table}
   \caption{Photometry for J041828}
   \centering
     {\small
       \begin{tabular}{ccccl}
         \linss
         $\lambda$
         &$S_\nu$
         &$\sigma_\mathrm{abs}$\tnmk{a}
         &Beam
         \\
         ($\mu$m)
         &(mJy)
         &(mJy)
         &(arcsec)
         &Instrument
         \\
         \linss
         3.4  &  0.260    &  0.012  &  2.3	  &  WISE\\
         3.6  &  0.406    &  0.010  &  1.7	  &  Spitzer/IRAC\\
         4.5  &  0.641    &  0.015  &  1.7	  &  Spitzer/IRAC\\
         4.6  &  0.515    &  0.024  &  2.9	  &  WISE\\
         5.8  &  0.796    &  0.045  &  1.9	  &  Spitzer/IRAC\\
         8.0  &  1.303    &  0.052  &  2.0	  &  Spitzer/IRAC\\
         12  &  1.36    &  0.17  &  7.6	  &  WISE\\
         22  &  3.33    &  1.04  &  13.8  &  WISE\\
         24  &  2.96    &  0.30  &  6.0	  &  Spitzer/MIPS\\
         70  &  20 	   &  5	  &  5.6	  &  Herschel/PACS\\
         160  &  $<52$    &  $-$	  &  11	  &  Herschel/PACS\\
         250  &  $<160$    &  $-$	  &  18	  &  Herschel/SPIRE\\
         350  &  $<95$    &  $-$	  &  25	  &  Herschel/SPIRE\\
         500  &  $<84$    &  $-$	  &  35	  &  Herschel/SPIRE\\
         1200  &  $<5.4$  &  $-$	  &  11	  &  IRAM30m/MAMBO\\
         13000 &$<0.045$  &$-$   &  $1.4\time1.2$ &JVLA\\
         36000 &$<0.123$  &$-$   &  $2.4\times 1.7$ &JVLA\\
         \linss
       \end{tabular}

       \begin{minipage}{8cm}
         \tntxt{a}{Absolute flux uncertainty.
           We adopted an absolute flux uncertainty of 25\% for \emph{Herschel}
           measurements.
         }
       \end{minipage}
     }
 \label{tab:app_sed_J041828}
 \end{table}

 \begin{table}
   \caption{Photometry for J041836}
   \centering
     {\small
       \begin{tabular}{ccccl}
         \linss
         $\lambda$
         &$S_\nu$
         &$\sigma_\mathrm{abs}$\tnmk{a}
         &Beam
         \\
         ($\mu$m)
         &(mJy)
         &(mJy)
         &(arcsec)
         &Instrument
         \\
         \linss
         1.23  &  0.248    &  0.041  &  2.5	  &  2MASS\\
         1.66  &  0.167    &  0.017  &  2.5	  &  2MASS\\
         2.16  &  0.506    &  0.057  &  2.5	  &  2MASS\\
         3.4  &  0.742    &  0.022  &  2.3	  &  WISE\\
         3.6  &  0.760    &  0.013  &  1.7	  &  Spitzer/IRAC\\
         4.5  &  1.015    &  0.017  &  1.7	  &  Spitzer/IRAC\\
         4.6  &  0.982    &  0.037  &  2.9	  &  WISE\\
         5.8  &  1.119    &  0.044  &  1.9	  &  Spitzer/IRAC\\
         8.0  &  2.577    &  0.061  &  2.0	  &  Spitzer/IRAC\\
         12  &  3.18    &  0.24  &  7.6	  &  WISE\\
         22  &  12.5    &  1.0	  &  13.8  &  WISE\\
         24  &  11.7  &  0.4	  &  6.0	  &  Spitzer/MIPS\\
         70  &  $<36$ 	 	&$-$	  &  5.6	  &  Herschel/PACS\\
         160  &  $<63$    &  $-$	  &  11	  &  Herschel/PACS\\
         250  &  $<131$    &  $-$	  &  18	  &  Herschel/SPIRE\\
         350  &  $<91$    &  $-$	  &  25	  &  Herschel/SPIRE\\
         500  &  $<101$    &  $-$	  &  35	  &  Herschel/SPIRE\\
         1200  &  $<7.2$  &  $-$	  &  11	  &  IRAM30m/MAMBO\\
         13000 &0.13   &0.03  &  $2.0\times 1.8$ &JVLA\\
         36000 &$<0.105$  &$-$   &  $2.4\times1.8$ &JVLA\\
         \linss
       \end{tabular}

       \begin{minipage}{8cm}
         \tntxt{a}{Absolute flux uncertainty.
           We adopted an absolute flux uncertainty of 25\% for \emph{Herschel}
           measurements.
         }
       \end{minipage}
     }
 \label{tab:app_sed_J041836}
 \end{table}

 \begin{table}
   \caption{Photometry for J041847}
   \centering
     {\small
       \begin{tabular}{ccccl}
         \linss
         $\lambda$
         &$S_\nu$
         &$\sigma_\mathrm{abs}$\tnmk{a}
         &Beam
         \\
         ($\mu$m)
         &(mJy)
         &(mJy)
         &(arcsec)
         &Instrument
         \\
         \linss
         3.4  &  0.528    &  0.020  &  2.3	  &  WISE\\
         3.6  &  0.609    &  0.014  &  1.7	  &  Spitzer/IRAC\\
         4.5  &  0.876    &  0.021  &  1.7	  &  Spitzer/IRAC\\
         4.6  &  0.725    &  0.028  &  2.9	  &  WISE\\
         5.8  &  1.020    &  0.049  &  1.9	  &  Spitzer/IRAC\\
         8.0  &  2.124    &  0.050  &  2.0	  &  Spitzer/IRAC\\
         12  &  3.30    &  0.25  &  7.6	  &  WISE\\
         22  &  9.96    &  1.10  &  13.8  &  WISE\\
         24  &  7.24    &  0.29  &  6.0	  &  Spitzer/MIPS\\
         70  &  $<25$ 	 	&$-$	  &  5.6	  &  Herschel/PACS\\
         160  &  165    &  41	  &  11	  &  Herschel/PACS\\
         250  &  87    &  22	  &  18	  &  Herschel/SPIRE\\
         350  &  $<46$    &  $-$	  &  25	  &  Herschel/SPIRE\\
         500  &  $<41$    &  $-$	  &  35	  &  Herschel/SPIRE\\
         1200  &  $<5.8$  &  $-$	  &  11	  &  IRAM30m/MAMBO\\
         13000 &$<0.045$  &$-$   &  $2.0\times1.8$ &JVLA\\
         36000 &$<0.123$  &$-$   &  $3.4\times2.8$ &JVLA\\
         \linss
       \end{tabular}

       \begin{minipage}{8cm}
         \tntxt{a}{Absolute flux uncertainty.
           We adopted an absolute flux uncertainty of 25\% for \emph{Herschel}
           measurements.
         }
       \end{minipage}
     }
 \label{tab:app_sed_J041847}
 \end{table}

 \begin{table}
   \caption{Photometry for J041913}
   \centering
     {\small
       \begin{tabular}{ccccl}
         \linss
         $\lambda$
         &$S_\nu$
         &$\sigma_\mathrm{abs}$\tnmk{a}
         &Beam
         \\
         ($\mu$m)
         &(mJy)
         &(mJy)
         &(arcsec)
         &Instrument
         \\
         \linss
         3.4  &  0.285    &  0.011  &  2.3	  &  WISE\\
         3.6  &  0.320    &  0.010  &  1.7	  &  Spitzer/IRAC\\
         4.5  &  0.456    &  0.014  &  1.7	  &  Spitzer/IRAC\\
         4.6  &  0.461    &  0.022  &  2.9	  &  WISE\\
         5.8  &  0.732    &  0.048  &  1.9	  &  Spitzer/IRAC\\
         8.0  &  1.429    &  0.045  &  2.0	  &  Spitzer/IRAC\\
         12  &  2.83    &  0.23  &  7.6	  &  WISE\\
         22  &  9.08    &  1.09  &  13.8  &  WISE\\
         24  &  7.10    &  0.28  &  6.0	  &  Spitzer/MIPS\\
         70  &  $<20$ 	 	&$-$	  &  5.6	  &  Herschel/PACS\\
         160  &  $<40$    &  $-$	  &  11	  &  Herschel/PACS\\
         250  &  $<89$    &  $-$	  &  18	  &  Herschel/SPIRE\\
         350  &  $<50$    &  $-$	  &  25	  &  Herschel/SPIRE\\
         500  &  $<50$    &  $-$	  &  35	  &  Herschel/SPIRE\\
         1200  &  $<5.9$  &  $-$	  &  11	  &  IRAM30m/MAMBO\\
         13000 &0.21   &0.02  &  $1.3\times 1.2$ &JVLA\\
         36000 &0.42   &0.05  &  $2.3\times 1.8$ &JVLA\\
         \linss
       \end{tabular}

       \begin{minipage}{7.7cm}
         \tntxt{a}{Absolute flux uncertainty.
           We adopted an absolute flux uncertainty of 25\% for \emph{Herschel}
           measurements.
         }
       \end{minipage}
     }
 \label{tab:app_sed_J041913}
 \end{table}

 \begin{table}
   \caption{Photometry for J041938}
   \centering
     {\small
       \begin{tabular}{ccccl}
         \linss
         $\lambda$
         &$S_\nu$
         &$\sigma_\mathrm{abs}$\tnmk{a}
         &Beam
         \\
         ($\mu$m)
         &(mJy)
         &(mJy)
         &(arcsec)
         &Instrument
         \\
         \linss
         1.23  &  0.161    &  0.034  &  2.5	  &  2MASS\\
         1.66  &  0.220    &  0.055  &  2.5	  &  2MASS\\
         2.16  &  0.366    &  0.058  &  2.5	  &  2MASS\\
         3.4  &  0.277    &  0.013  &  2.3	  &  WISE\\
         3.6  &  0.278    &  0.009  &  1.7	  &  Spitzer/IRAC\\
         4.5  &  0.315    &  0.012  &  1.7	  &  Spitzer/IRAC\\
         4.6  &  0.299    &  0.022  &  2.9	  &  WISE\\
         5.8  &  0.300    &  0.036  &  1.9	  &  Spitzer/IRAC\\
         8.0  &  1.303    &  0.063  &  2.0	  &  Spitzer/IRAC\\
         12  &  1.97    &  0.22  &  7.6	  &  WISE\\
         22  &  6.1    &  1.4	  &  13.8  &  WISE\\
         24  &  6.66  &  0.32  &  6.0	  &  Spitzer/MIPS\\
         70  &  93 	   &  23	  &  5.6	  &  Herschel/PACS\\
         160  &  157    &  39	  &  11	  &  Herschel/PACS\\
         250  &  134    &  33	  &  18	  &  Herschel/SPIRE\\
         350  &  68    &  17	  &  25	  &  Herschel/SPIRE\\
         500  &  67    &  17	  &  35	  &  Herschel/SPIRE\\
         1200  &  $<5.0$  &  $-$	  &  11	  &  IRAM30m/MAMBO\\
         13000 &0.11   &0.03  &  $2.0\times 1.8$ &JVLA\\
         36000 &0.11   &0.03  &  $3.3\times 2.7$ &JVLA\\
         \linss
       \end{tabular}

       \begin{minipage}{7.7cm}
         \tntxt{a}{Absolute flux uncertainty.
           We adopted an absolute flux uncertainty of 25\% for \emph{Herschel}
           measurements.
         }
       \end{minipage}
     }
 \label{tab:app_sed_J041938}
 \end{table}

 \begin{table}
   \caption{Photometry for J042019}
   \centering
     {\small
       \begin{tabular}{ccccl}
         \linss
         $\lambda$
         &$S_\nu$
         &$\sigma_\mathrm{abs}$\tnmk{a}
         &Beam
         \\
         ($\mu$m)
         &(mJy)
         &(mJy)
         &(arcsec)
         &Instrument
         \\
         \linss
         1.23  &  0.233    &  0.034  &  2.5	  &  2MASS\\
         1.66  &  0.239    &  0.055  &  2.5	  &  2MASS\\
         2.16  &  0.308    &  0.054  &  2.5	  &  2MASS\\
         3.4  &  0.296    &  0.011  &  2.3	  &  WISE\\
         3.6  &  0.377    &  0.009  &  1.7	  &  Spitzer/IRAC\\
         4.5  &  0.606    &  0.014  &  1.7	  &  Spitzer/IRAC\\
         4.6  &  0.570    &  0.027  &  2.9	  &  WISE\\
         5.8  &  1.150    &  0.046  &  1.9	  &  Spitzer/IRAC\\
         8.0  &  1.991    &  0.062  &  2.0	  &  Spitzer/IRAC\\
         12  &  2.70    &  0.23  &  7.6	  &  WISE\\
         22  &  3.7    &  1.2	  &  13.8  &  WISE\\
         24  &  4.32  &  0.29  &  6.0	  &  Spitzer/MIPS\\
         70  &  $<29$ 	 	&$-$	  &  5.6	  &  Herschel/PACS\\
         160  &  $<40$    &  $-$	  &  11	  &  Herschel/PACS\\
         250  &  $<98$    &  $-$	  &  18	  &  Herschel/SPIRE\\
         350  &  $<61$    &  $-$	  &  25	  &  Herschel/SPIRE\\
         500  &  $<65$    &  $-$	  &  35	  &  Herschel/SPIRE\\
         1200  &  $<5.4$  &  $-$	  &  11	  &  IRAM30m/MAMBO\\
         13000 &$<0.041$  &$-$   &  $1.3\times 1.2$ &JVLA\\
         36000 &$<0.205$  &$-$   &  $2.2\times 1.7$ &JVLA\\
         \linss
       \end{tabular}

       \begin{minipage}{8cm}
         \tntxt{a}{Absolute flux uncertainty.
           We adopted an absolute flux uncertainty of 25\% for \emph{Herschel}
           measurements.
         }
       \end{minipage}
     }
 \label{tab:app_sed_J042019}
 \end{table}

 \begin{table}
   \caption{Photometry for J042118 (after \citealt{Palau12}).}
   \centering
     {\small
       \begin{tabular}{ccccl}
         \linss
         $\lambda$
         &$S_\nu$
         &$\sigma_\mathrm{abs}$\tnmk{a}
         &Beam
         \\
         ($\mu$m)
         &(mJy)
         &(mJy)
         &(arcsec)
         &Instrument
         \\
         \linss
         0.89  &  $<0.0142$	&$-$	  &  1.	  &  SDSS\\
         1.03  &  $<0.0130$  	&$-$	  &  0.6	  &  UKIDSS/WFCAM\\
         1.25  &  0.0165    &  0.0003  &  0.6	  &  UKIDSS/WFCAM\\
         2.20  &  0.0453    &  0.0009  &  0.6	  &  UKIDSS/WFCAM\\
         3.4  &  0.196    &  0.010  &  2.3	  &  WISE\\
         3.6  &  0.249    &  0.008  &  1.7	  &  Spitzer/IRAC\\
         4.5  &  0.439    &  0.014  &  1.7	  &  Spitzer/IRAC\\
         4.6  &  0.409    &  0.030  &  2.9	  &  WISE\\
         5.8  &  0.674    &  0.045  &  1.9	  &  Spitzer/IRAC\\
         8.0  &  1.094    &  0.063  &  2.0	  &  Spitzer/IRAC\\
         12  &  1.06    &  0.28  &  7.6	  &  WISE\\
         22  &  3.79    &  0.38  &  13.8  &  WISE\\
         24  &  2.73    &  0.28  &  6.0	  &  Spitzer/MIPS\\
         70  &  $<21$ 	 	&$-$	  &  5.6	  &  Herschel/PACS\\
         160  &  $<43$    &  $-$	  &  11	  &  Herschel/PACS\\
         250  &  54    &  13	  &  18	  &  Herschel/SPIRE\\
         350  &  29    &  7	  &  25	  &  Herschel/SPIRE\\
         350  &  46    &  9	  &  10	  &  CSO/SHARC\\
         500  &  $<44$    &  $-$	  &  35	  &  Herschel/SPIRE\\
         1200  &  $<4.0$  &  $-$	  &  11	  &  IRAM30m/MAMBO\\
         13000	&$<0.036$ 	&$-$	  &  $1.3\times 1.2$	&JVLA\\
         36000	&$<0.132$ 	&$-$	  &  $2.2\times 1.7$	&JVLA\\
         \linss
       \end{tabular}

       \begin{minipage}{8cm}
         \tntxt{a}{Absolute flux uncertainty.
           We adopted an absolute flux uncertainty of 25\% for \emph{Herschel}
           measurements.
         }
       \end{minipage}
     }
 \label{tab:app_sed_J042118}
 \end{table}

 \begin{table}
   \caption{Photometry for J042123 (radio-galaxy)}
   \centering
     {\small
       \begin{tabular}{ccccl}
         \linss
         $\lambda$
         &$S_\nu$
         &$\sigma_\mathrm{abs}$\tnmk{a}
         &Beam
         \\
         ($\mu$m)
         &(mJy)
         &(mJy)
         &(arcsec)
         &Instrument
         \\
         \linss
         1.23  &  0.287    &  0.035  &  2.5	  &  2MASS\\
         1.66  &  0.490    &  0.059  &  2.5	  &  2MASS\\
         2.16  &  0.602    &  0.062  &  2.5	  &  2MASS\\
         3.4  &  0.618    &  0.018  &  2.3	  &  WISE\\
         3.6  &  0.609    &  0.014  &  1.7	  &  Spitzer/IRAC\\
         4.5  &  0.709    &  0.017  &  1.7	  &  Spitzer/IRAC\\
         4.6  &  0.576    &  0.043  &  2.9	  &  WISE\\
         5.8  &  0.825    &  0.047  &  1.9	  &  Spitzer/IRAC\\
         8.0  &  1.510    &  0.060  &  2.0	  &  Spitzer/IRAC\\
         12  &  1.95    &  0.24  &  7.6	  &  WISE\\
         22  &  4.90    &  0.49  &  13.8  &  WISE\\
         24  &  4.65    &  0.27  &  6.0	  &  Spitzer/MIPS\\
         70  &  18 	   &  5	  &  5.6	  &  Herschel/PACS\\
         160  &  $<34$    &  $-$	  &  11	  &  Herschel/PACS\\
         250  &  $<53$    &  $-$	  &  18	  &  Herschel/SPIRE\\
         350  &  $<31$    &  $-$	  &  25	  &  Herschel/SPIRE\\
         500  &  $<28$    &  $-$	  &  35	  &  Herschel/SPIRE\\
         1200  &  $<5.6$  &  $-$	  &  11	  &  IRAM30m/MAMBO\\
         13000 &0.48   &0.02  &  $1.3\times1.2$ &JVLA\\
         36000 &0.84   &0.07  &  $2.2\times 1.8$ &JVLA\\
         \linss
       \end{tabular}

       \begin{minipage}{7.7cm}
         \tntxt{a}{Absolute flux uncertainty.
           We adopted an absolute flux uncertainty of 25\% for \emph{Herschel}
           measurements.
          }
       \end{minipage}
     }
 \label{tab:app_sed_J042123}
 \end{table}

 \begin{acronym}
   \acro{BD}{brown dwarf}
   \acro{CASA}{Common Astronomy Software Application}
   \acro{CSO}{Caltech Submillimeter Observatory}
   \acro{IMF}{initial mass function}
   \acro{JVLA}{Jansky Very Large Array}
   \acro{NRAO}{National Radio Astronomy Observatory}
   \acro{SED}{spectral energy distribution}
   \acro{VeLLO}{Very Low Luminosity Object}
   \acro{YSO}{young stellar object}
 \end{acronym}

\end{document}